\documentclass[aps,prb,twocolumn,groupedaddress]{revtex4}
\usepackage{graphicx}
\usepackage{amssymb}
\usepackage{amsmath}
\usepackage{bm}
\usepackage{xcolor}

\newcommand{\ket}[1]{|{#1}\rangle}
\newcommand{\bra}[1]{\langle{#1}|}
\newcommand{\braket}[2]{\langle{#1}|{#2}\rangle}
\newcommand{\abs}[1]{\left|{#1}\right|}

\newcommand{\Heff}{H_\textrm{eff}}

\newcommand{\rrr}{\vec{r}}
\newcommand{\DDD}{\mathcal{D}}

\newcommand{\ua}{\ensuremath{\uparrow}}
\newcommand{\da}{\ensuremath{\downarrow}}

\begin{document}

\title{Edge-state enhanced transport in a 2-dimensional quantum walk}
\author{Janos K. Asboth}
\affiliation{
Institute for Solid State Physics and Optics, 
Wigner Research Centre, Hungarian Academy of Sciences, 
H-1525 Budapest P.O. Box 49, Hungary}
\author{Jonathan M. Edge}
\affiliation{Nordita, KTH Royal Institute of Technology and Stockholm University, Roslagstullsbacken 23
106 91 Stockholm, Sweden}
\begin{abstract}
Quantum walks on translation invariant regular graphs spread
quadratically faster than their classical counterparts. The same
coherence that gives them this quantum speedup inhibits, or even stops
their spread in the presence of disorder. We ask how to create an
efficient transport channel from a fixed source site (A) to fixed
target site (B) in a disordered 2-dimensional discrete-time quantum
walk by cutting some of the links. We show that the somewhat
counterintuitive strategy of cutting links along a single line
connecting A to B creates such a channel.  The efficient transport
along the cut is due to topologically protected chiral edge states,
which exist even though the bulk Chern number in this system
vanishes. We give a realization of the walk as a periodically driven
lattice Hamiltonian, and identify the bulk topological invariant
responsible for the edge states as the quasienergy winding of this
Hamiltonian. 
\end{abstract}
\pacs{05.30.Rt,03.67.-a,03.65.Vf}
\maketitle

\section{Introduction}
\label{sec:introduction}

The Discrete-Time Quantum Walk (DTQW, or quantum walk for
short)\cite{kempe_2003}, a quantum mechanical generalization of the
random walk, has in the recent years received more and more attention
from both the theoretical and experimental side.  The main drive to
understand the properties of the DTQW come from its possible use for
quantum information processing, be it quantum search
algorithms\cite{Shenvi03}, or even general purpose quantum
computation\cite{dtqw_universal}. Experiments on quantum walks range
from realizations on trapped
ions\cite{travaglione_02,roos_ions,schmitz_ion}, to cold atoms in
optical lattices\cite{meschede_science,alberti_electric_experiment},
and on light on an optical
table\cite{gabris_prl,schreiber_science,peruzzo_science_2010,
  white_photon_prl,sciarrino_twoparticle}, but there are many
other experimental proposals\cite{rydberg_walk,kalman_09}.

The distinguishing feature of quantum walks is that on regular graphs,
they spread faster than their classical counterparts: the
root-mean-square distance of the walker from the origin increases with
the number $N$ of steps as $\mathcal{O}(N)$, rather than
$\mathcal{O}(\sqrt{N})$ as in the classical case.  This can be put to good
use for algorithms based on quantum walks\cite{Shenvi03} that find a
marked state among $N$ states in only $\mathcal{O}(\sqrt{N})$ steps,
outperforming their classical counterparts -- the same scaling
advantage as of the Grover algorithm\cite{grover_prl}, which can also
be understood as a DTQW.  The intuitive explanation for this ballistic
scaling is that a DTQW can be seen as a stroboscopic simulator for an
effective Hamiltonian, and thus, in a clean system, its eigenstates
are plane waves.

If we understand a DTQW to be a stroboscopic simulator for a
Hamiltonian, we can expect that static disorder can impede the
spreading of the walk, even bringing it to a complete standstill,
through Anderson localization\cite{tiggelen_phystoday}.
This prediction has been mathematically proven for some types of
one-dimensional DTQWs\cite{joye_10,ahlbrecht_2011}, and even observed
in an optical implementation \cite{gabris_anderson}. However, even in
one dimension, some types of disorder lead to a slow, subdiffusive
spreading of the walk rather than complete
localization\cite{obuse_delocalization}; this phenomenon can also be
explained in terms of the effective
Hamiltonian\cite{obuse_delocalization,brouwer_delocalization}. Two-dimensional
DTQWs are also expected to suffer Anderson
localization\cite{Svozilik2012}, although in some cases disorder causes
diffusion\cite{jonathan_2014}.

In this paper we address the question: is there a way to create an
efficient transport channel in a 2-dimensional split-step DTQW (2DQW)
that defeats localization even if static disorder is present? We take
a DTQW on a square lattice, with two special sites: $A$, where the
walk is started from, and $B$, where we want the walker to ultimately
end up, rather than escaping to infinity or remaining in the vicinity
of $A$. To create a channel, we cut links on the lattice, thus
restricting the movement of the walker. The first idea, cutting out a
narrow island, with $A$ on the one end, and $B$ on the other, is
rendered ineffective by static disorder. We find a somewhat
counterintuitive strategy that does work, however: cutting the links
along a single line connecting $A$ to $B$ creates a conveyor belt for
the walker, transporting it efficiently and ballistically from $A$ to
$B$ even in the presence of considerable amount of static disorder.

The way that a cut along a line on the lattice of the quantum walk
forms a robust conveyor belt for the walker is reminiscent of how
electrons are transported along line defects by edge states in
topological insulators\cite{rmp_kane}. This seems to be a promising
direction for an understanding of the transport mechanism, since the
effective Hamiltonians of DTQWs can be engineered to realize all
classes of topological phases in 1 and 2
dimensions\cite{kitagawa_exploring}. However, the effective
Hamiltonian of the 2DQW is topologically
trivial\cite{kitagawa_exploring}. Thus, if there is a bulk topological
invariant protecting these states from disorder, it is not covered by
standard theory\cite{schnyder_tenfold}.

The topological structure of DTQWs is in fact richer than that of
time-independent Hamiltonians, and exploration of that structure is
far from complete. The telltale signs of extra topology are protected
edge states at the edges of bulks where the topological invariants of
the effective Hamiltonian predict none. An example is one-dimensional
DTQWs with chiral symmetry, where such edge states have been detected
in an optical experiment\cite{kitagawa_observation}, and have been
predicted to exist between two bulks with the same effective
Hamiltonian\cite{asboth_prb}. In that case, the extra topological
structure responsible for the protection of these states has been
found, and can be described based on time-delayed effective
Hamiltonians\cite{asboth_2013}, scattering
matrices\cite{scattering_walk2014}, or as winding numbers of one part
of the timestep operator\cite{asboth_2014}.  Edge states between two
bulks in the 2DQW have been found
numerically\cite{kitagawa_introduction}, but the extra topological
invariants that they indicate are unknown.

In this paper we show that there are chiral (one-way propagating)
edge states along a cut in a 2DQW, and identify the bulk topological
invariant responsible for their appearance. We map the quantum walk to
a periodically driven Hamiltonian, and thus identify the invariant as
the winding number found by Rudner et al.\cite{rudner_driven}, which
we refer to as Rudner invariant.

The paper is structured as follows. We introduce the type of 2DQW we
consider, together with the prescription of how to cut links on the
graph, in Section \ref{sec:definitions}. Then, in Section
\ref{sec:first_transport}, we consider two strategies to enhance
transport in the 2DQW: in a clean case, the straightforward, ``island
cut'' approach works fine, but in the presence of disorder, only the
less intuitive, ``line cut'' approach gives efficient transport.  We
show that there are edge states along the line cut in Section
\ref{sec:edge_states}. In Section \ref{sec:top_invariants} we find the
bulk topological invariants responsible for the edge states.  In
Sect.~\ref{sec:robustn-conv-belt} we consider the effects of disorder
on the edge state transport.

\section{Definitions }
\label{sec:definitions}

Of the wide variety of two-dimensional quantum walks, we choose the
split-step walk on a square lattice (2DQW), defined in
Ref.~\onlinecite{kitagawa_exploring}, for its simplicity: it requires
only two internal states for the walker. In
this section we recall the definition of the 2DQW,
introduce the conditional wavefunction method which allows us to treat
transport in the quantum walk setting, discuss how to cut links in the
quantum walk and how disorder is introduced.

\begin{figure}
\includegraphics[width=6cm]{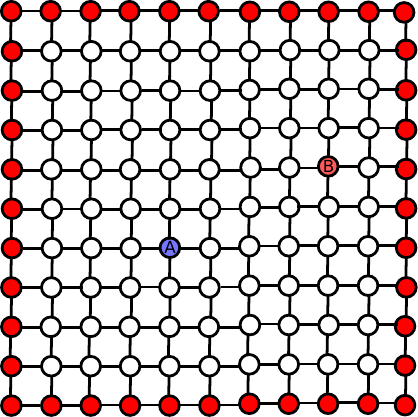}
\caption{Layout of the 2-dimensional quantum walk, with a source at
  $A$, a detector at the target site at $B$, and detectors at the
  edges. For the conditional wavefunction, the detectors play the role
  of absorbers.
}
\label{fig:alice_bob}
\end{figure}

\subsection{Walker wavefunction and time evolution operator}
\label{sec:walk-wavef-time}

We consider a particle, or \emph{walker}, on a square lattice, with
two internal states, which we refer to as spin. The wavefunction of
the walker can be written as
\begin{align} 
\ket{\Psi} &= \sum_{\rrr\in \DDD} (
\Psi(\rrr,\uparrow)\ket{\rrr,\uparrow} + 
\Psi(\rrr,\downarrow)\ket{\rrr,\downarrow}).
\end{align} 
Here $\rrr=(x,y)$ is a 2-dimensional vector of integers, which labels
the nodes of the lattice, taken from
$\DDD=\{(x,y)|x=1,\ldots,x_\text{max}, y = 1,\ldots,y_\text{max}\}$.
The walker is initialized at site $A=(x_A,y_A)$ as
\begin{align}
  \ket{\Psi(t=0)} &= \ket{A,\uparrow}.
\end{align}
The dynamics of the walker takes place in discrete time $t\in \mathbb{N}$, 
and is determined by 
\begin{align} 
\ket{\Psi(t+1)} &= U \ket{\Psi(t)};\\ 
U &= S_y R_2 S_x R_1.
\label{eq:U_def}
\end{align}
The operator $R_j$, with $j=1,2$, denotes a rotation of the spin about
the $y$ axis, 
\begin{align} 
R_j &= \sum_{\rrr\in \DDD} \ket{\rrr}\bra{\rrr} \otimes
e^{-i\theta_j(\rrr)\sigma_y}.
\end{align} 
The angles $\theta_1$ and $\theta_2$ of the first and second
rotation can depend on the position $\rrr=(x,y)$ of the walker.
The operators $S_x$ and $S_y$ denote spin-dependent translations along
links between the sites on the lattice,
\begin{align} 
S_x &= \sum_{\rrr\in \DDD} \ket{\rrr+\hat{x},\uparrow}\bra{\rrr,\uparrow} + 
\ket{\rrr,\downarrow}\bra{\rrr+\hat{x},\downarrow};\\
S_y &= \sum_{\rrr\in \DDD} \ket{\rrr+\hat{y},\uparrow}\bra{\rrr,\uparrow} + 
\ket{\rrr,\downarrow}\bra{\rrr+\hat{y},\downarrow},
\end{align} 
where $\hat{x}=(1,0)$ and $\hat{y}=(0,1)$.
 

\subsection{Conditional wavefunction} 

We want to measure how efficient transport is to a given site, $B =
(x_B,y_B)$, as opposed to propagation to the boundary of the system,
denoted by the sites $C_j$, as shown in Fig.~\ref{fig:alice_bob}.  We
place a detector at site $B$, and at the boundary sites $C_j$. After
every timestep, each detector performs a dichotomic measurement on the
wavefunction: if the walker is at the detector, it is detected, if
not, it is undisturbed. To calculate the resulting probability
distribution for the transmission times, we compute the
\emph{conditional wavefunction} $\ket{\Psi(t)}$, conditioned on no
detection events up to time $t$. To obtain the time evolution of the
conditional wavefunction, at the end of each timestep the components
of the wavefuntion at the sites $B$ and $C_j$ are projected out,
\begin{align}
\Psi(t) &= \Big(1-\ket{B}\bra{B}-\sum_j\ket{C_j}\bra{C_j} \Big) 
U \ket{\Psi(t-1)}.
\label{eq:timestep_def}
\end{align}
Note that measurements are performed at each step, but since the
measurement record is kept, the whole process is still completely
coherent.

The norm of the walker's wavefunction, $\braket{\Psi(t)}{\Psi(t)}$, is
the probability that the particle is still in the system after $t$
steps.  Due to the postselection involved in the timestep,
Eq.~\eqref{eq:timestep_def}, this norm decreases over time as the
walker is found at $B$ (successful transmission) or leaks out at the
edges (transmission failure).
The probability of success, i.e., of detecting the walker at $B$ at
time $t$, is given by
\begin{align}
p_t &= \sum_{s=\uparrow,\downarrow} 
\abs{\bra{B,s} U \ket{\Psi(t-1)}}^2.
\label{eq:def_pt}
\end{align}
The arrival probability at time $t$ is the summed probabilities of absorption up to time $t$ and is given by
\begin{align}
  \label{eq:def_of_arrival_prob_Pt}
  P_t=\sum_{t'=1}^t p_{t'}
\end{align}

\subsection{Disorder through the rotation angles. }
We will consider the effects of disorder that enters the system
through the angles $\theta$.  The rotation angles become position
dependent, uncorrelated random variables, chosen from a uniform
distribution,
\begin{align}
\theta_j(\vec r) \in [\theta_j-\delta, \theta_j + \delta].
\label{eq:angle_disorder}
\end{align}
In this paper we will consider 
time-independent (i.e.,
static, or quenched) disorder, i.e., the angles $\theta$ depend only on
position, but not on time. 
The effects of disorder will be addressed in
section~\ref{sec:robustn-conv-belt}.



\subsection{Cutting links}  
To enhance transport, we consider modifying the graph on which the
walk takes place by cutting some of the links.  If the link between
sites $(x,y)$ and $(x+1,y)$ is cut, the $\ua$ component of the
wavefunction is not transported from site $(x,y)$ to $(x+1,y)$ during
the $S_x$ shift operation and similarly the $\da$ component from
$(x+1,y)$ is not shifted to $(x,y)$. The analogous definition for cut
links holds for the $S_y$ operation between sites $(x,y)$ and
$(x,y+1)$.

If
we were dealing with a lattice Hamiltonian instead of a lattice
timestep operator, cutting a link could be done by just setting the
corresponding hopping amplitude to 0. In the case of the timestep
operator, however, maintaining the unitary of the time evolution --
orthogonal states always have to stay orthogonal\cite{asboth_prb} --
is more involved. 
The only sensible unitary and short-range way to do that is to induce
a spin flip instead of a hop, with possibly an additional phase
factor.  This extra phase plays an important role in the 1D quantum
walk, where it affects the quasienergy of the end
states\cite{asboth_prb}. For 2D quantum walks, however, this extra
phase factor unimportant. For convenience, we flip the spin
using $-i\sigma_y$.

The complete shift operator $S_d$, with $d=x$ or $y$, including the
prescription for cutting the links, reads
\begin{align}
  S_d&=\sum_{\vec r \in \mathcal{L}_d}
  \left(
    \ket{\vec r + \hat d, \ua} \bra{\vec r,\ua} + \ket{\vec r,
      \da}\bra{\vec r + \hat d, \da}
  \right)
  \nonumber\\&\quad
  + \sum_{\vec r \in \mathcal{C}_d}
  \left(
    \ket{\vec r, \da}\bra{\vec r , \ua} - \ket{\vec r + \hat d, \ua}\bra{\vec r + \hat d, \da}
  \right).
  \label{shift_op_x_with_cut}
\end{align}
Here $\mathcal{L}_d$ is the set of vectors $\vec r$ such that the link
between node at $\vec r$ and the node at $\vec r + \hat d$ is not cut,
while its complement $\mathcal{C}_d$ is the set of vectors to nodes
$\vec r$ for which the link connecting them to node $\vec r + \hat d$
has been cut, with $\hat d$ denoting the unit vector in the direction
$d$ (i.e., $\hat{x}$ or $\hat{y}$).


\section{Transport in the presence of a cut}
\label{sec:first_transport}
We now address the question: which links should we cut to optimize the
transport from A to B?  The first idea that comes to mind to ensure
efficient transport is to cut out a narrow island from the lattice: at
the one end of the island is $A$, the source, at the other end $B$,
the site where we want the walker to be transported to. However, as we
see, in the presence of disorder, there is a much more efficient
construction.

\subsection{The island cut}

\begin{figure}
\includegraphics[width=6cm]{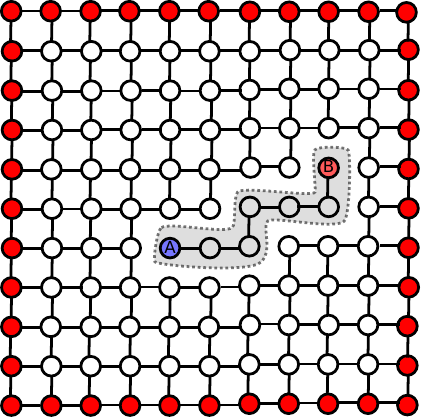}
\caption{
To increase the efficiency of transport from $A$ to $B$, the first
idea is to cut an island that will form a transport channel, as
indicated by the dashed line. All links crossing the dashed line are
cut; a particle attempting to hop across a cut link will have its spin
flipped instead of hopping.  }
\label{fig:alice_bob_island}
\end{figure}

\begin{figure}
\includegraphics[width=4.4cm]{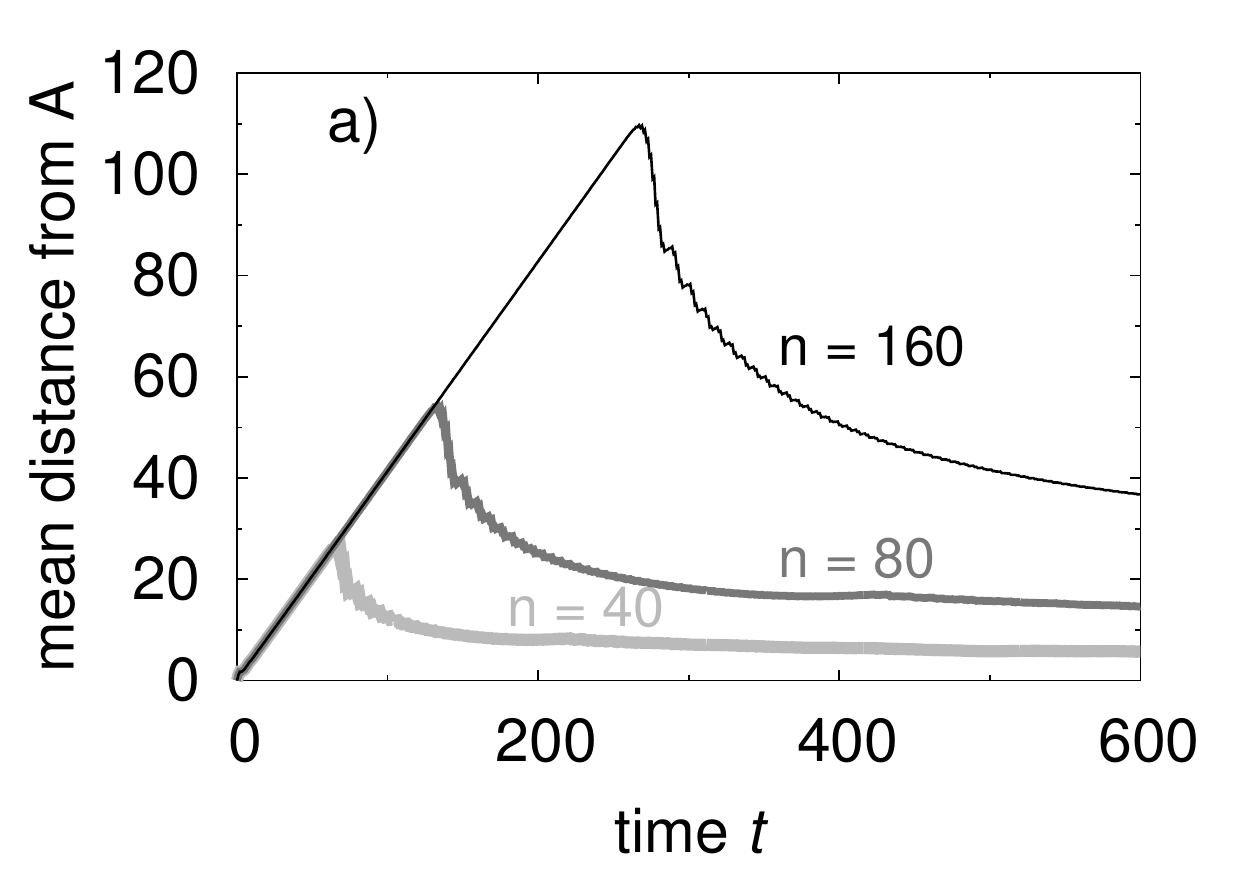}%
\includegraphics[width=4.4cm]{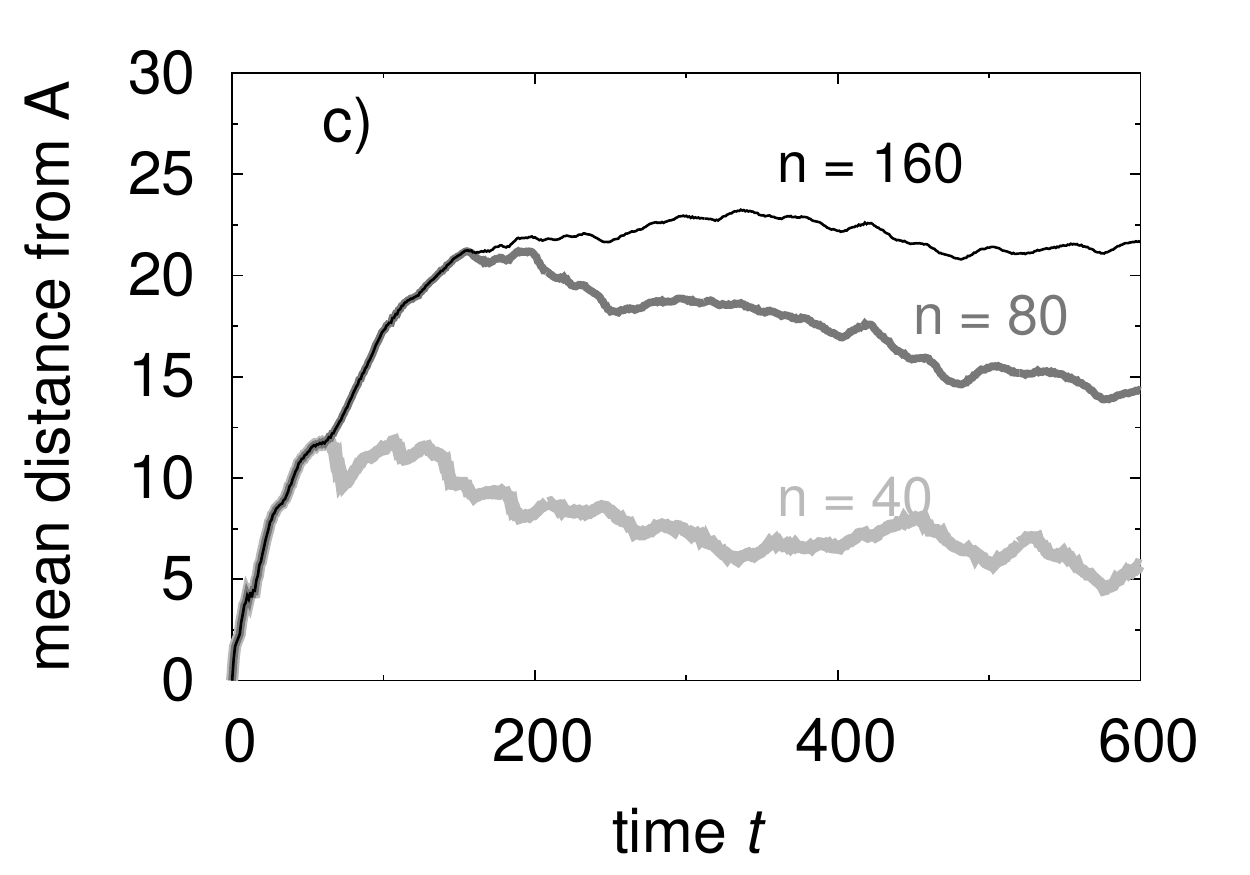}
\includegraphics[width=4.4cm]{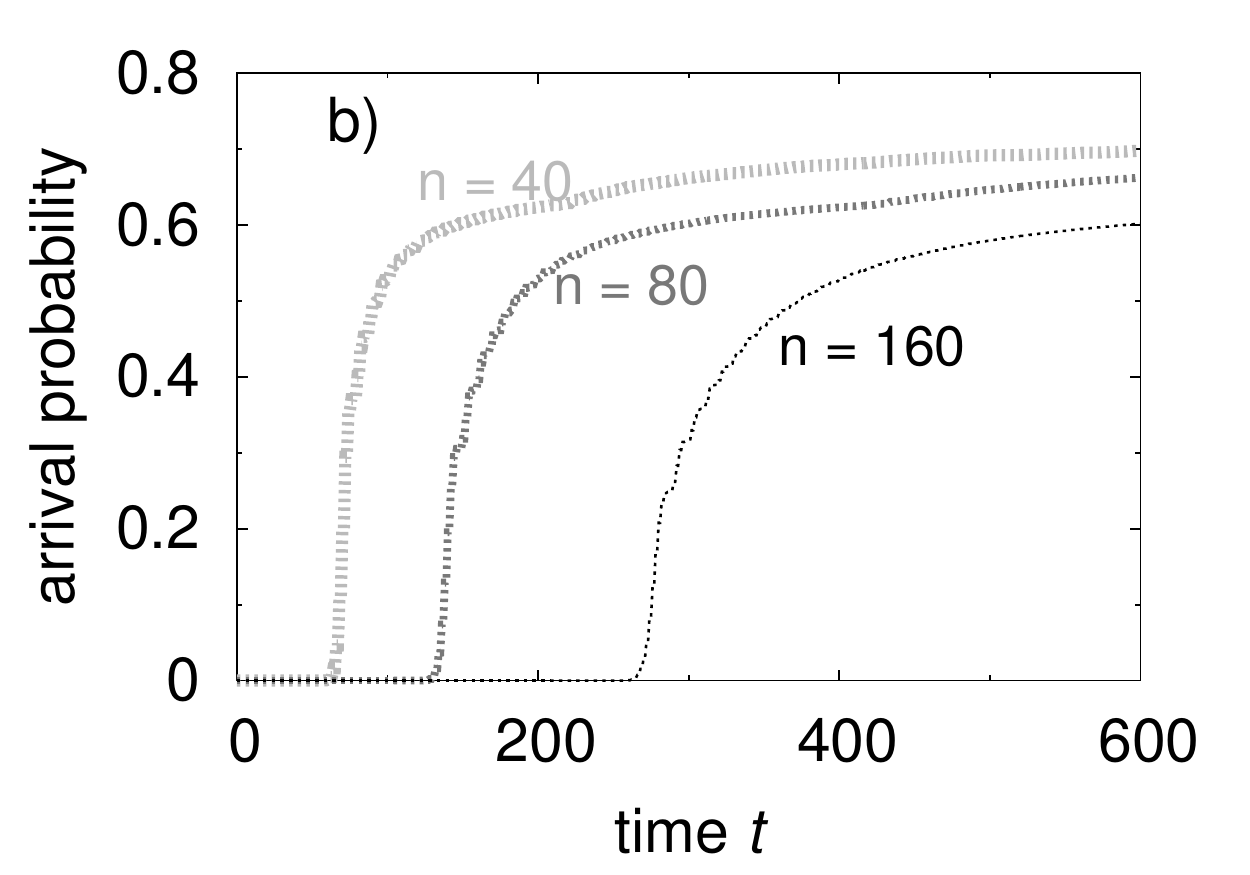}%
\includegraphics[width=4.4cm]{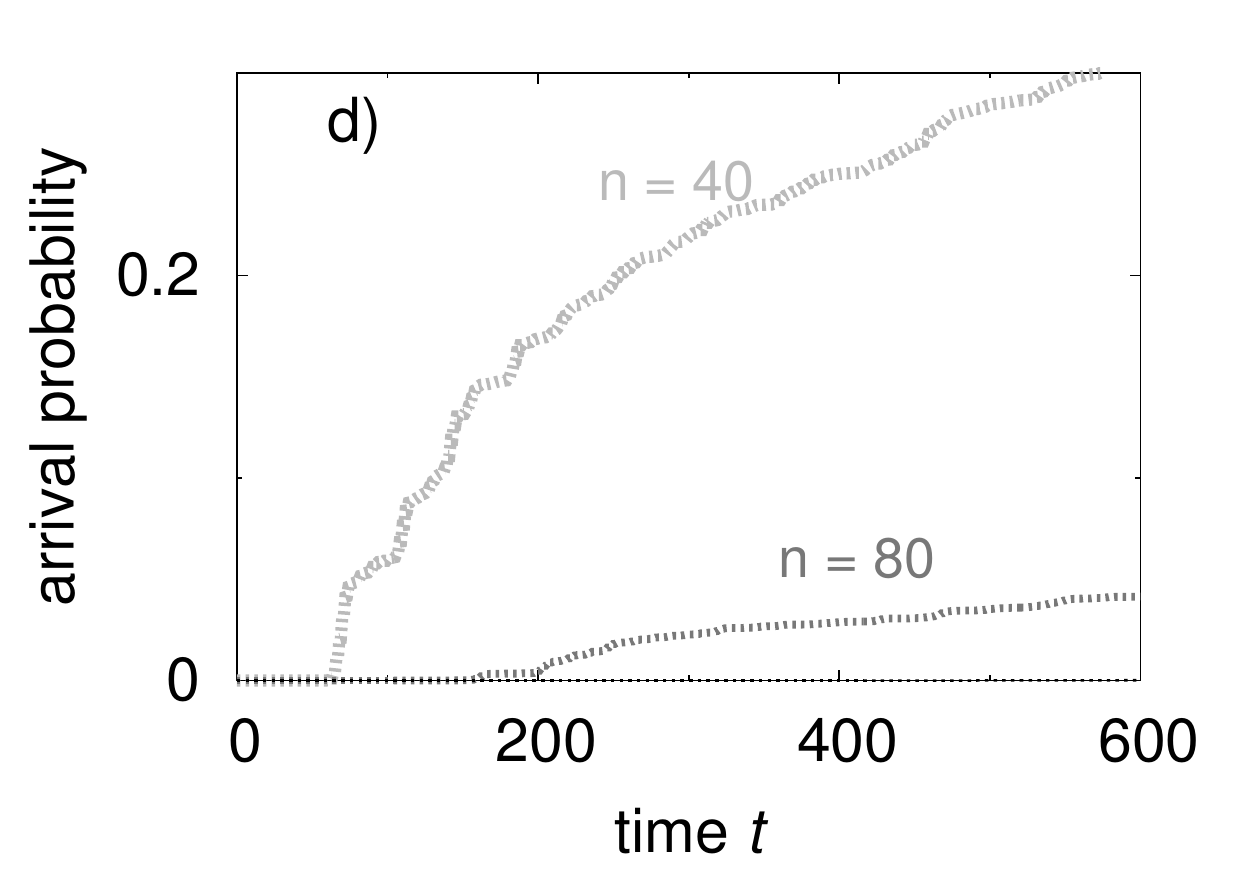}
\caption{
 Mean displacement of the quantum walker (continuous lines)
  and transmission probability (dotted lines) in the geometry of the
  ``island'' of Fig.~\ref{fig:alice_bob_island}, for a horizontal
  island of fixed width of 1, and varying island length of 40 (thick
  light gray), 80 (medium gray), and 160 (thin black).  Mean rotation
  angles are set to $\theta_1=0.35\pi$, $\theta_2 =0.15\pi$. Without
  disorder, $\delta=0$, the wavefunction spreads ballistically (a) and
  the transmission probability reaches a value close to 1 as the
  wavepacket arrives (b). To illustrate the effects of disorder, we
  set $\delta=0.2 \pi$, and use a single disorder realization, varying
  only the distance $n$ between $A$ and $B$ (and correspondingly, the
  length of the island). For large enough system ($n=160$), the mean
  distance from $A$ saturates at around 30 (c), and in this case there
  is virtually no transmission ((d): $P_t<10^{-4}$ for $n=160$ for all
  times $t$).}
\label{fig:localization_barrier}
\end{figure}

Perhaps the most straightforward way to ensure that the walker gets
from $A$ to $B$ is to restrict its motion to a narrow island
connecting these two sites, by cutting links as illustrated in
Fig.~\ref{fig:alice_bob_island}.  In a clean system, this strategy
achieves the desired effect. Simulations on large system sizes, shown
in Fig.~\ref{fig:localization_barrier}.a, show a high success
probability, independent of system size (island length), with a time
required for transport proportional to the length of the island,
indicating ballistic transport. 

The simple strategy of cutting out an island to guide the walker to
$B$ no longer works if there is quenched disorder in the rotation
angles.  As shown in Fig.~\ref{fig:localization_barrier}b), the time
evolution of the walker's wavefunction now shows signs of
localization. With a disorder of $\delta\theta = 0.07 \pi$, the
average distance from the origin stops growing after some time,
independent of system size. 

\subsection{The single line cut}

\begin{figure}
\includegraphics[width=6cm]{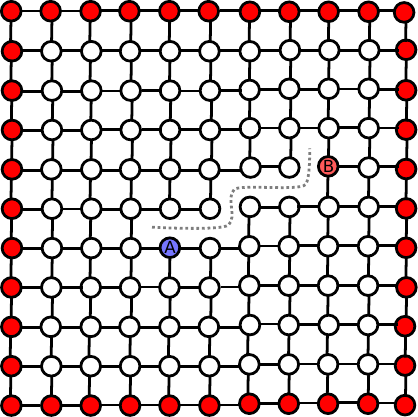}%
\caption{An alternative strategy to create a channel between $A$ and
  $B$ is a single cut.}
\label{fig:alice_bob_1cut}
\end{figure}

There is a somewhat counterintuitive strategy to defeat localization,
and ensure efficient transport from $A$ to $B$ even with static
disorder. This involves cutting links along a line from A to B, as
shown in Fig.~\ref{fig:alice_bob_1cut}.

\begin{figure}
\includegraphics[width=4.4cm]{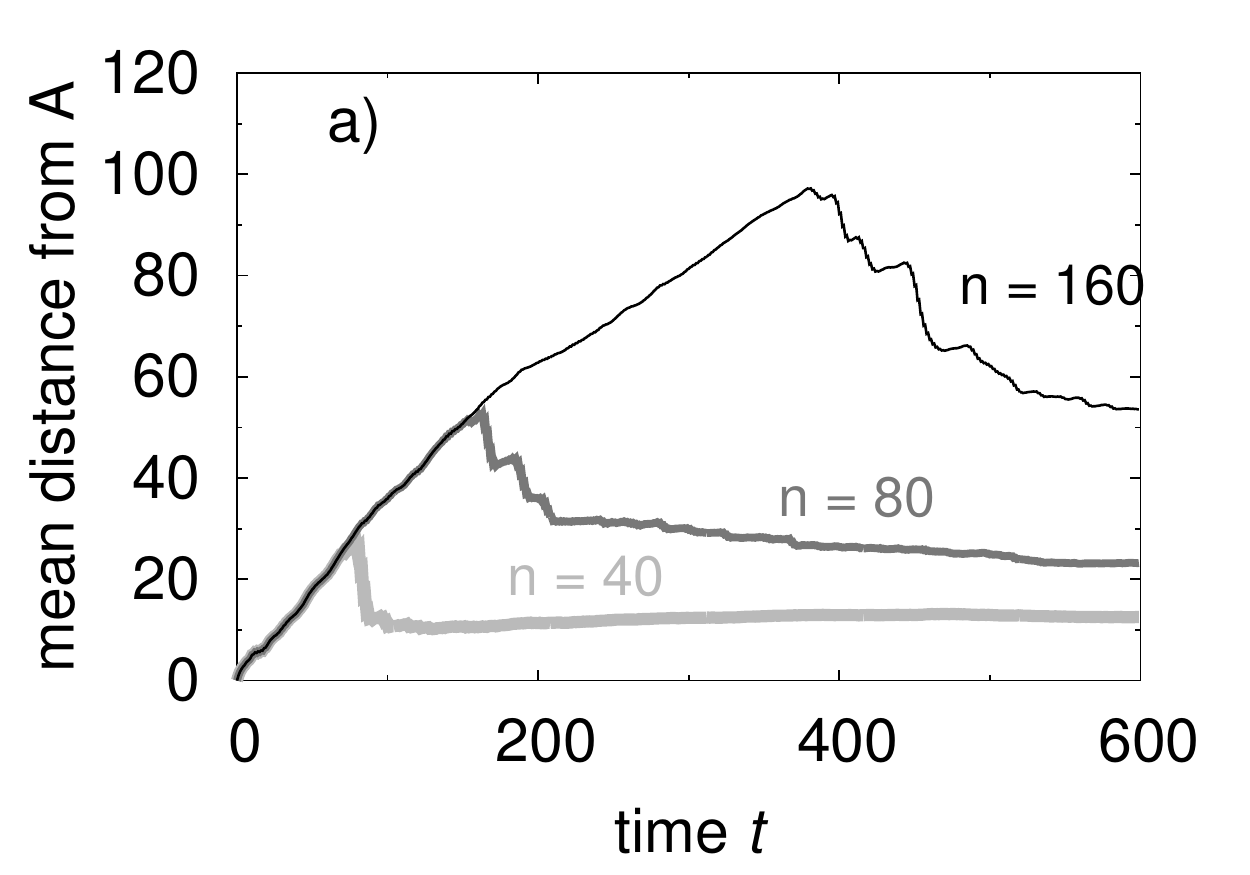}%
\includegraphics[width=4.4cm]{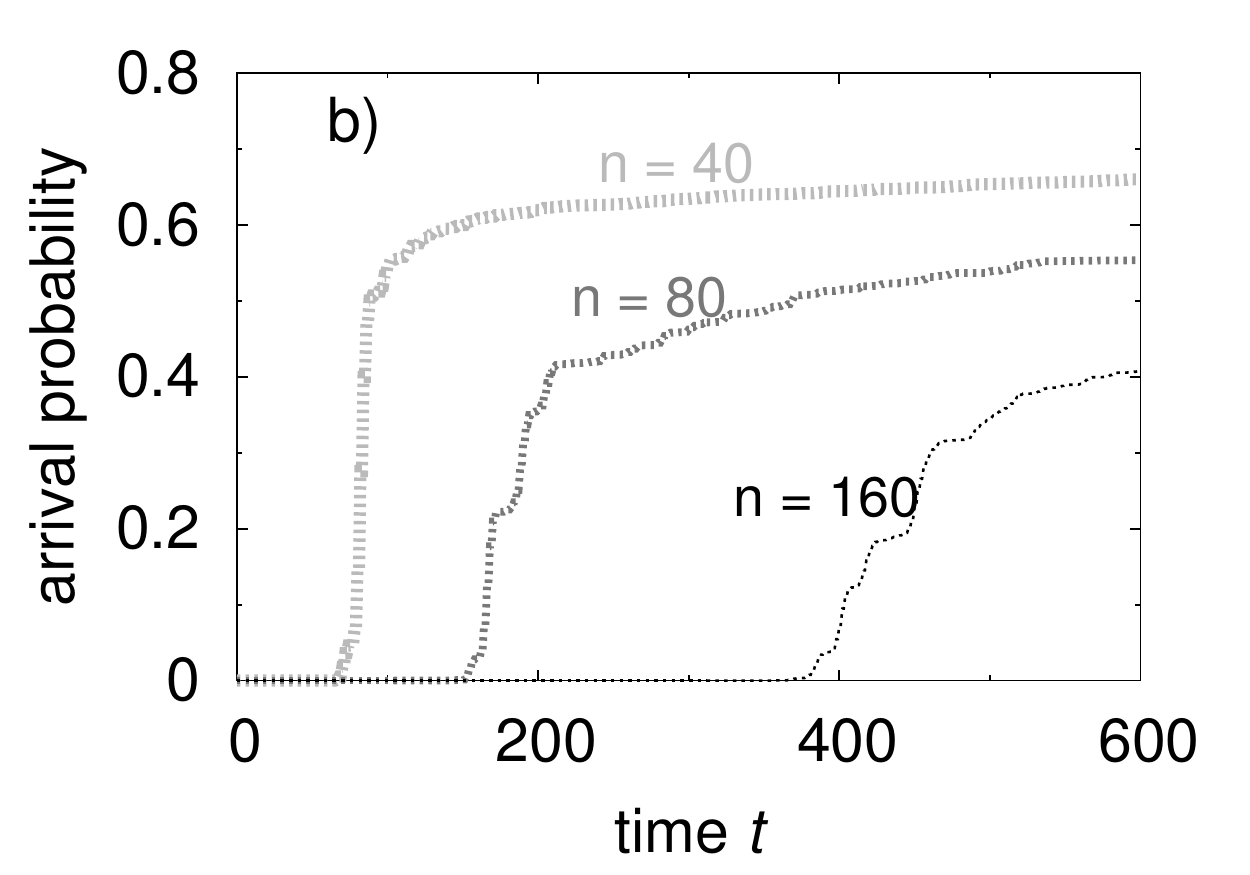}
\caption{ Mean displacement of the quantum walker (a, continuous
  lines) and transmission probability (b, dotted lines) in the
  geometry of the ``single cut'' of Fig.~\ref{fig:alice_bob_1cut}, for
  a horizontal cut.  A single realization of a disordered quantum walk
  is taken, with mean rotation angles $\theta_1=0.35\pi$, $\theta_2
  =0.15\pi$, and disorder $\delta=0.3 \pi$. The $A-B$ distance $n$ is
  varied: $n=40$ (thick light gray), $n=80$ (medium gray), and $n=160$
  (thin black). The walker propagates ballistically along the cut (b),
  and arrives at $B$ with a high probability (c).}
\label{fig:transport_with_cut}
\end{figure}

As shown in Fig.~\ref{fig:transport_with_cut}, in spite of the
disorder, the single cut ensures ballistic propagation of the quantum
walker and greatly enhances the transmission probability: the line of
cut links acts like a conveyor belt for the quantum walker.
Although for the detailed numerics we used cuts that are along a
straight line, numerical examples convincingly show that the shape of
the cut can delay the transport, but not inhibit it.  For an example,
see the Appendix \ref{app:star}.

The rest of this paper is devoted to this conveyor belt mechanism.
Our principal aims will be to answer the following two questions: Why
does the conveyor mechanism work? How robust is it?

\section{Edge states along a cut}
\label{sec:edge_states}

In this section we show that the single cut transports the walker
efficiently from the source $A$ to the target site $B$ because the
quantum walk has unidirectional (chiral) edge states along the cut. We
find the edge states along the cut using the effective Hamiltonian.


%





The effective Hamiltonian $\Heff$ of a DTQW is defined as 
\begin{align}
\Heff &= i \log U, 
\label{eq:def_heff}
\end{align}
where $U$, as in Eq.~\eqref{eq:U_def}, is the unitary timestep
operator of the quantum walk without the projectors corresponding to
the measurements. We fix the branch cut of the logarithm to be along
the negative part of the real axis. If we only look at the DTQW at
integer times $t$, we cannot distinguish a DTQW from the time
evolution that would be produced by the time-independent lattice
Hamiltonian $\Heff$, since,
\begin{align}
\ket{\Psi(t)} &= U^t \ket{\Psi(0)}= e^{-i \Heff t} \ket{\Psi(0)}
\,\,\text{ for } t \in \mathbb{N}.
\end{align}
Every DTQW is thus a stroboscopic simulator for its effective
Hamiltonian $\Heff$.

\begin{figure}
\includegraphics[width=0.8\columnwidth]{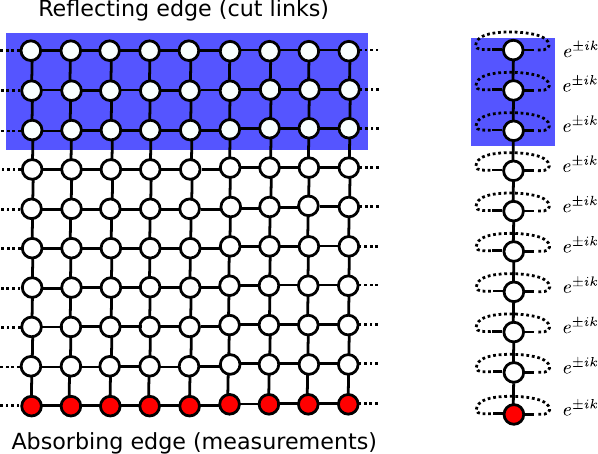}
\caption{For the analytical calculation, we consider a simple geometry
  with reflecting edge on top, and absorbers on the bottom. An
  infinite strip (left) can be treated as a 1-dimensional
  chain with twisted boundary conditions, i.e. with periodic
  boundaries along $x$ with an extra phase of $e^{\pm ikx}$ for
  right/left hopping. The top three rows, with dark (blue) background
 are defined as the edge region.}
\label{fig:edge_state_def}
\end{figure}

We now consider the quasienergy dispersion relation of a clean system
in the vicinity of (below) a horizontal cut, as shown in
Fig.~\ref{fig:edge_state_def}. We make use of translation
invariance, and use $k$ to denote the quasimomentum along $x$, a
conserved quantity. We take system of width $1$ ($x=1$) and height $L$
($y=1,\ldots,L$), with modified periodic boundary conditions along
both directions. Along $x$, twisted boundary conditions are taken,
i.e., periodic boundary conditions with an extra phase factor of
$e^{\mp ik}$ for right/left shifts, with $k$ denoting the
quasimomentum we are interested in. Along $y$, we leave the periodic
boundary conditions, but cut the link connecting site $(1,L)$ with
$(1,1)$, and we insert an absorber at $(1,1)$. We diagonalize the
timestep operator $U$ on this system, obtaining the eigenvalues
$\lambda_n = \abs{\lambda_n} e^{-i\varepsilon_n}$ and the
corresponding eigenvectors $\ket{\Psi}_n$. The magnitudes
$\abs{\lambda_n} \le 1$ give us information about the lifetime of the
states, while the phases $\varepsilon_n$ can be identified with the
quasienergies. Repeating this procedure for $-\pi < k \le \pi $ gives
us the dispersion relation of a clean strip with a cut at the top and
absorbers at the bottom.

\begin{figure}
\includegraphics[width=0.95\columnwidth]{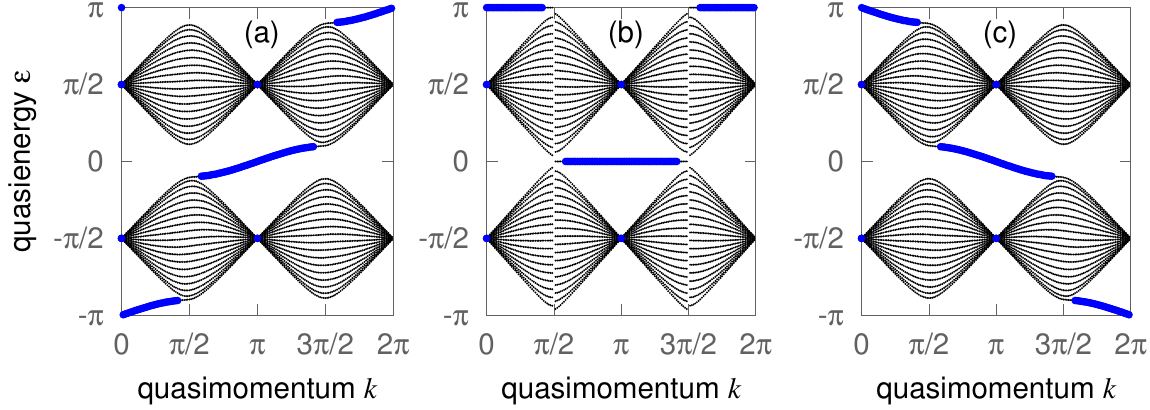}
\includegraphics[width=0.95\columnwidth]{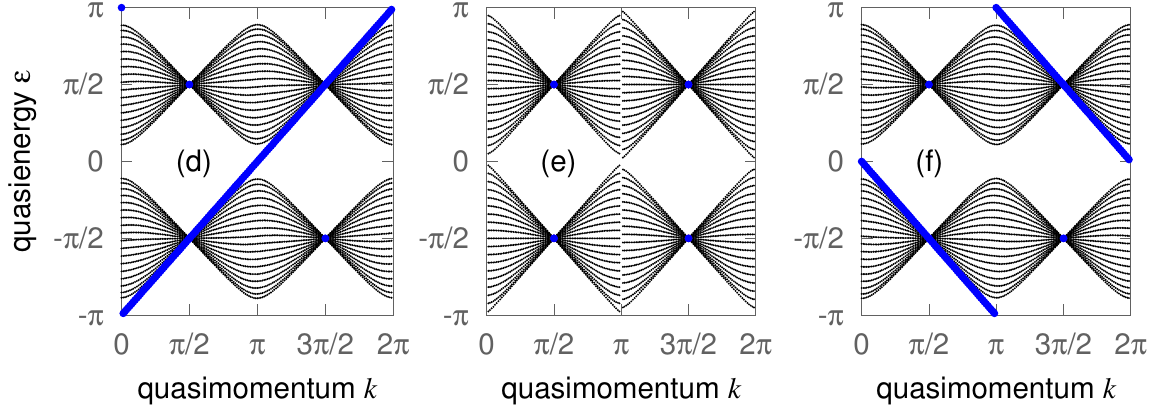}
\caption{Dispersion relation of a 2DQW on a strip with cut links
  on top, and absorbers at the bottom. Quasienergies of long-lived
  states (magnitude of Floquet eigenvalue higher than $0.9$) are shown,
  with edge states (more than 80\% of the weight on the top three
  rows) highlighted in thick (blue). The bulk gap is closed and reopened
  by setting the rotation angles to 
(a): $\theta_1=0.35\pi$,
  $\theta_2=0.15\pi$, (b): $\theta_1=\theta_2=0.25\pi$, (c): 
  $\theta_1=0.15\pi$, $\theta_2=0.35\pi$. 
(d): $\theta_1=0.65\pi$,
  $\theta_2=0.15\pi$, (e): $\theta_1=0.75\pi$, $\theta_2=0.25\pi$, (f): 
  $\theta_1=0.85\pi$, $\theta_2=0.35\pi$. 
}
\label{fig:stripe_dispersions}
\end{figure}

We show the numerically obtained dispersion relation of the 2DQW on a
stripe with an edge in Fig.~\ref{fig:stripe_dispersions}. 
We omitted states with short lifetimes, whose eigenvalue of $U$ has
magnitude $\abs{\lambda}<0.9$. We used thick (blue) to highlight edge
states, defined as states for which $\abs{\braket{L}{\Psi}}^2+
\abs{\braket{L-1}{\Psi}}^2 + \abs{\braket{L-2}{\Psi}}^2 > 0.9$.
Whenever the gaps around $\varepsilon=0$ and $\varepsilon=\pi$ are
open, one can clearly see edge states traversing these gaps. The edge
states are unidirectional (i.e., chiral), and propagate in the same
direction in the two gaps.

\begin{figure}
\includegraphics[width=5cm]{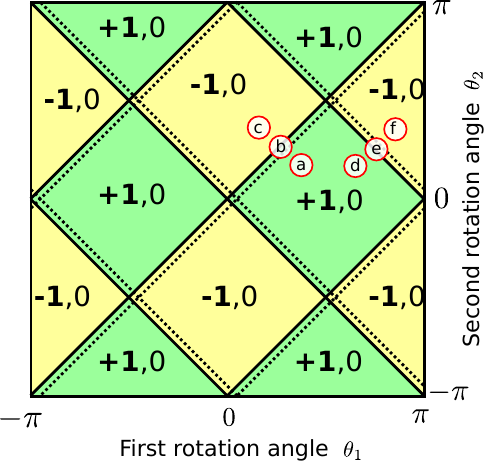}
\caption{Parameter space of the split-step 2D discrete-time quantum
  walk. Along continuous (dotted) lines, the bulk quasienergy gap
  around 0 ($\pi$) quasienergy closes. Each gapped domain supports
  edge states near a cut, at both quasienergies 0 and $\pi$. The
  number of these edge states (equal to the Rudner winding number as
  per Eq.~\eqref{eq:def_rudner_winding}), is written in bold. The Chern
  number, which is always 0 due to the sublattice symmetry, is shown in
  normal typeface.  }
\label{fig:phase_space_windings}
\end{figure}

We obtained simple analytical formulas for the dispersion relations of
the edge states along the horizontal cut, for $\varepsilon\approx 0$
and $\varepsilon\approx \pi$, using the transfer matrix method. We
relegate the details to the Appendix \ref{app:edge_state}, and
summarize the main results here. When $\sin(\theta_1+\theta_2) > 0$,
the edge states are around $k = \varepsilon = 0$ and
$k=\varepsilon=\pm\pi$ (as in Fig.~\ref{fig:stripe_dispersions}a-d),
when $\sin(\theta_1 + \theta_2) < 0$, they are around $k=\pm\pi,
\varepsilon=0$ and $k=0, \varepsilon=\pm\pi$ (as in
Fig.~\ref{fig:stripe_dispersions}f).  Near the center of the gaps, the
edge states group velocity reads
\begin{align}
v=\frac{d \varepsilon}{d k} &= 
\sin(\theta_2-\theta_1) \text{ sign }\left[\sin(\theta_1+\theta_2) \right].
\label{eq:group_velocity}
\end{align}
The edge states decay exponentially towards the bulk as 
$\Psi \propto e^{-\abs{y}/\xi}$, where $y$ is the distance from the edge. 
Using the analytical
calculations of Appendix \ref{app:edge_state}, we obtain the
penetration depth $\xi$ of the edge states into the bulk as
\begin{align}
\xi &=-\left( \text{log } 
\frac{1-\abs{\sin(\theta_1+\theta_2}}{\abs{\cos(\theta_1-\theta_2)}} 
\right)^{-1}.
\label{eq:edge_penetration}
\end{align} 
Although the penetration depth and the magnitude of the group velocity
can depend on the orientation of the edge, the direction of
propagation of these chiral edge states constitutes a topological
invariant. We show this topologically protected quantity as a function
of the parameters $\theta_1$ and $\theta_2$ by boldface numbers in
Fig.~\ref{fig:phase_space_windings}.

The direction of propagation (chirality) of the edge states is
topologically protected: it can only be changed if the rotation angles
$\theta_j$ are themselves changed so that the system is taken across a
gap closing point. There are two different scenarios here,
corresponding to gap closings where $\theta_1-\theta_2=n\pi$ (lines
slanting upwards in Fig.~\ref{fig:phase_space_windings}, e.g., labels
(a)-(c) in Figs.~\ref{fig:stripe_dispersions} and
\ref{fig:phase_space_windings}), and where $\theta_1+\theta_2=n\pi$
(lines slanting downwards in Fig.~\ref{fig:phase_space_windings},
e.g., labels (d)-(f) in Figs.~\ref{fig:stripe_dispersions} and
\ref{fig:phase_space_windings}).  In the first case, during the gap
closing, the number of edge states constituting the edge mode does not
change, their penetration depth, Eq.~\eqref{eq:edge_penetration} stays
finite, it is only the edge mode velocity that goes to zero and then
changes sign, see Eq.~\eqref{eq:group_velocity}.  In the second case,
the velocity of the edge mode does not change as the gap is closed; it
is the number of edge states that goes to zero and then grows
again. In this case, the penetration depth $\xi$ diverges as the gap
is closed.  The two scenarios of this paragraph correspond to edge
states at a zigzag or an armchair edge in the Haldane
model\cite{haldane_model} (e.g., Fig.~5.~of
Ref.~\onlinecite{PhysRevB.89.205408}).

\section{Topological invariant of the 2-dimensional split-step quantum
walk }
\label{sec:top_invariants}

In a free lattice system with unitary dynamics, the number of
unidirectional (chiral) edge states in the bulk energy gap cannot be
altered by any local changes in the dynamics, as long as the bulk
energy gap is open. Thus, the number of such edge states constitutes a
topological invariant for each bulk gap.  In time-independent lattice
Hamiltonians, this invariant can be obtained from the bulk Hamiltonian
as the sum of the Chern numbers of all the bands with energy below the
gap.  The Chern number for the bands of the 2DQW, however, is always
zero, due to a discrete sublattice symmetry of the timestep operator,
as we show in Appendix \ref{app:sublattice_symmetry}.  Thus, there has
to be some other bulk topological invariant of the 2DQW. This extra
topological invariant is also indicated by the fact that edge states
appear at an interface between two domains of the 2DQW with the same
Chern number\cite{kitagawa_introduction}.  We will now identify this
bulk topological invariant.

\subsection{The Rudner Invariant in periodically driven quantum systems}


A candidate for the topological invariant of the 2DQW is
the winding number of periodically driven 2-dimensional lattice
Hamiltonians found by Rudner et al.\cite{rudner_driven}, which we
summarize here. Consider a periodically driven lattice Hamiltonian,
\begin{align}
H(t+1, k_x,k_y) &= H(t,k_x,k_y).
\end{align}
The unitary time evolution operator for one complete period reads 
\begin{align}
U(k_x,k_y) &= \mathbb{T} e^{-i\int_0^1 H(k_x,k_y,t) dt}. 
\end{align}
Next, define a loop in the following way, 
\begin{align}
U_2(t, k_x,k_y) = \left\{ \begin{array}{rl}
 \mathbb{T} e^{-2i\int_0^t H(k_x,k_y,2t') dt'} &\mbox{ if $t<\frac{1}{2}$} \\
 e^{2i(t-1/2) \Heff} U(k_x,k_y) 
&\mbox{ if $t\ge\frac{1}{2}$}
       \end{array} \right.
\end{align}
This corresponds to going forward in time until $t=1/2$ with the full
Hamiltonian, and then backwards in time with the effective
Hamiltonian, as in Eq.~\eqref{eq:def_heff}, whose branch cut is chosen
at $\varepsilon=\pi$. Thus, $U_2(t=0)=U_2(t=1)=1$, and $U_2(t=1/2)=U$.

The winding number associated with $U_2$ is
\begin{align}
W[U_2] &= \frac{1}{8 \pi^2} \int dt dk_x dk_y \text{Tr } \big(
U_2^{-1} \partial_t U_2 \cdot \nonumber \\
\quad &\quad [U_2^{-1}\partial_{k_x} U_2,
  U_2^{-1}\partial_{k_y} U_2] \big).
\label{eq:def_rudner_winding}
\end{align}
As Rudner et al.\cite{rudner_driven} show, the periodically driven
system will have a number $W$ of chiral edge states in addition to
those predicted by the Chern numbers of the bands. These edge states
appear in each gap, including the gap around $\varepsilon=\pi$ (if
there is a gap there; if not, the branch cut of the logarithm in
Eq.~\eqref{eq:def_heff} needs to be shifted to be in a gap).

\subsection{Rudner invariant from an equivalent  lattice Hamiltonian }

Rudner's invariant is defined for periodically driven lattice
Hamiltonians, not DTQWs. To define this invariant for the 2DQW, we
need a realization of the 2DQW as time periodic Hamiltonian.  We
construct such a realization analogously to the one-dimensional
case\cite{asboth_tarasinski_delplace}.

\begin{figure}
\includegraphics[width=3cm]{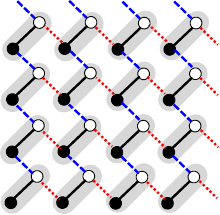}
\hspace{0.5cm}
\includegraphics[width=4.5cm]{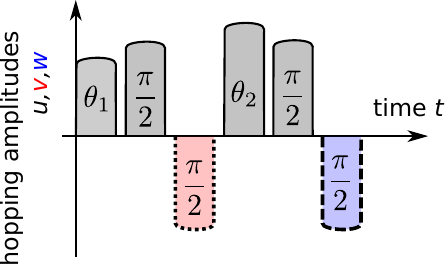}
\caption{Left: the lattice on which the 2DQW is realized as a continuously
  driven Hamiltonian. Gray shaded unit cells include two sites
  each. The three types of hoppings allowed are intracell (black),
  horizontal intercell (red) and vertical intercell (blue). Right: the drive
  sequence for the lattice Hamiltonian.  }
\label{fig:driven_H}
\end{figure}

We consider a square lattice of unit cells, each containing two sites,
denoted by filled circles $\bullet$ and empty circles $\circ$, as
shown in Fig.~\ref{fig:driven_H}. These sites are identified with states of the walker as 
\begin{align}
c^\dagger_{x,y,\bullet} \ket{0} &= \ket{x,y,\uparrow};& 
c^\dagger_{x,y,\circ} \ket{0} &= -i \ket{x,y,\downarrow};.
\end{align}
We take a nearest neighbor hopping Hamiltonian on this lattice, 
without any onsite terms, 
\begin{align}
H(t) &= \sum_{x,y} \big(
u(t) \hat{c}_{x,y,\bullet}^\dagger 
\hat{c}_{x,y,\circ}
\,+\,v(t) \hat{c}_{x,y,\bullet}^\dagger 
\hat{c}_{x-1,y,\circ} \nonumber \\
\quad &\quad \,\,+\,\,w(t) \hat{c}_{x,y,\bullet}^\dagger 
\hat{c}_{x,y-1,\circ}
\,+ h.c.
\,\big).
\label{eq:ssh_walk} 
\end{align}
We distinguish between three kinds of hoppings. \emph{Intracell}
hoppings, along the black lines in the grey unit cells in
Fig.~\ref{fig:driven_H}, have amplitudes
$u(t)$. \emph{Horizontal intercell} hoppings, along the dotted red
lines in in Fig.~\ref{fig:driven_H}, have amplitudes
$v(t)$. Finally, \emph{vertical intercell} hoppings, along the dashed
blue lines in Fig.~\ref{fig:driven_H}, have amplitudes
$w(t)$. 

To realize the 2DQW, we use a non-overlapping sequence of pulses where
at any time, only one type of hopping is switched on. 
A pulse of intracell
hopping $u$ of area $\pi/2$, followed by a pulse of intercell hopping
$v$, of area $-\pi/2$, realizes the operation $S_x$; if the pulse of
$u$ is followed by a pulse of $w$ of area $-\pi/2$, we obtain
$S_y$. The pulse sequence realizing a timestep of the 2DQW then
consists of 6 pulses, shown in Fig.~\ref{fig:driven_H}, and summarized
using the Heaviside function $\chi(x)=(\text{sign}(x)+1)/2$ as
\begin{align}
G(t) &= 6 \chi\left(t+\frac{1}{12}\right)\chi\left(\frac{1}{12}-t\right) ;\\
u(t) &=
\theta_1 G\Big(t-\frac{1}{12}\Big) + \frac{\pi}{2} G\Big(t-\frac{3}{12}\Big) \nonumber \\
\quad &\quad+ \theta_2 G\Big(t-\frac{7}{12}\Big) + \frac{\pi}{2}
G\Big(t-\frac{9}{12}\Big);\\
v(t) &= -\frac{\pi}{2} G\Big(t-\frac{5}{12}\Big);\\
w(t) &= -\frac{\pi}{2} G\Big(t-\frac{11}{12}\Big);
\end{align}

For this continuously driven Hamiltonian, we calculate the Rudner
invariant numerically, discretizing the integral of
Eq.~\eqref{eq:def_rudner_winding}, and find quantized values to a
great precision. The results are shown in
Fig.~\ref{fig:phase_space_windings}. We checked numerically that these
invariants correctly predict the edge states at reflective edges, and
also reproduce the edge states between different bulk phases of
Ref.~\onlinecite{kitagawa_introduction}.

\subsection{Cut links as a bulk phase: the 4-step 2D discrete-time quantum walk}

\begin{figure}
\includegraphics[width=2.5cm]{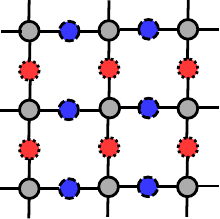}
\includegraphics[width=6cm]{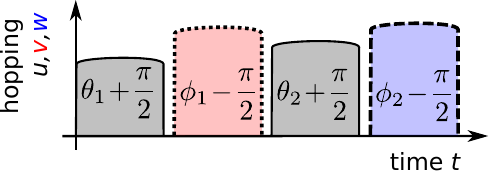}
\caption{The 4-step quantum walk is set on a Lieb lattice (left). The
  driving sequence of the corresponding continuously driven
  Hamiltonian consists of nonoverlapping pulses of arbitrary area
  (right). }
\label{fig:Hamiltonian_partial}
\end{figure}

To obtain a more complete picture of the conveyor belt mechanism, it
is instructive to view the line where the links are cut as the
limiting case of a long thin domain of a more general quantum walk
with modified parameters. To obtain this more general quantum walk, we
start from the continuous-time periodically driven Hamiltonian,
Eq.~\eqref{eq:ssh_walk}. There is a straightforward way to cut the link
in the $x$ $(y)$ direction: simply omit the pulse of $v(t)$ $(w(t)) $
from the sequence. This leads us to consider periodically driven
systems composed of pulses of arbitrary area, as represented in
Fig.~\ref{fig:Hamiltonian_partial},
\begin{align}
u(t) &=
\left(\theta_1 \!+ \frac{\pi}{2} \right) 
G\Big(t-\frac{1}{8}\Big) 
+
\left(\theta_2 \!+ \frac{\pi}{2} \right) 
G\Big(t-\frac{5}{8}\Big);\\
v(t) &= 
\left(\phi_1 - \frac{\pi}{2} \right) 
G\Big(t-\frac{3}{8}\Big);\\
w(t) &= 
\left(\phi_2 - \frac{\pi}{2} \right) 
G\Big(t-\frac{7}{8}\Big).
\end{align}

We can interpret this pulse sequence as a continuous-time 
realization of a  discrete-time quantum walk. This is the 4-step
walk, defined by  
\begin{align}
U = S_y\,e^{-i \phi_2 \sigma_y } \,S_y\,e^{-i \theta_2 \sigma_y } \, 
S_x\, e^{-i \phi_1 \sigma_y } \, S_x\, e^{-i \theta_1 \sigma_y }.  
\label{eq:4step_2D}
\end{align}
This walk is easiest represented on a Lieb lattice, as shown in
Fig.~\ref{fig:Hamiltonian_partial}. At the beginning and end of each
cycle, the walker is on one of the (gray) lattice sites with
coordination number 4, while during the timestep, it can also occupy
the (red and blue) sites with coordination number 2. 

\begin{figure}
\includegraphics[width=6.5cm]{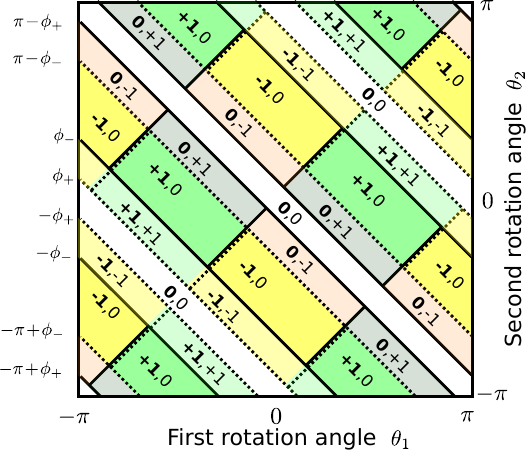}
\caption{Parameter space of the 4-step 2DQW as defined in
  Eq.~\eqref{eq:4step_2D}. Gapped domains, with Rudner winding numbers
  $W$ (boldface) and Chern numbers (normal typeface), are separated by
  lines, along which the bulk quasienergy gap around $\varepsilon=0$
  (continuous lines) or around $\varepsilon=\pi$ (dotted lines)
  closes.  Since sublattice symmetry of the walk is broken by the
  extra rotations through angles $\phi_1, \phi_2$, the gaps can close
  independently, and the Chern number can take on nonzero values.  The
  angles shown on the left are $\phi_+ = \abs{\phi_2+\phi_1}$,
  $\phi_-= \abs{\phi_2-\phi_1}$, assuming both of these are less than
  $\pi$. In the example shown, $\phi_1=-\pi/10$ and $\phi_2 = \pi/5$.}
\label{fig:phase_space_partial}
\end{figure}

The 4-step walk has two topological invariants: the Chern number $C$,
and the Rudner winding number $W$. Its Chern number can be nonzero,
because at the end of the timestep the walker can also return to its
starting point, and so it does not have the sublattice property
detailed in the Appendix \ref{app:sublattice_symmetry}. We find that,
depending on the angles $\phi_1,\phi_2,\theta_1,\theta_2$, the
invariants can take on the values $-1,0,+1$, as shown in
Fig.~\ref{fig:phase_space_partial}. In particular, the trivial
insulator, with $C=W=0$, is realized in the areas in parameter space
defined by $n \pi - \abs{\phi_1-\phi_2} < \theta_1-\theta_2 < n \pi +
\abs{\phi_1-\phi_2} $, for $n=0$ (including $U=-1$) and $n=\pm 1$
(including $U=1$). The phase with all links cut corresponds to
$\theta_1=\theta_2 = -\phi_1 = -\phi_2= -\pi/2$; in this case, the
time evolution operator does nothing to the state.

\section{Robustness of the conveyor belt in the presence of disorder}
\label{sec:robustn-conv-belt}
We now investigate how the transport along the cut is affected by static
disorder in the rotation angles $\theta_1 $ and
$\theta_2$, as defined in Eq.~\eqref{eq:angle_disorder}. 

\subsection{Effects of static disorder}
\label{sec:effects-stat-disord}

\begin{figure}[tb]
  \centering
  \def \picwidth {\columnwidth}
  \includegraphics[width=\picwidth]{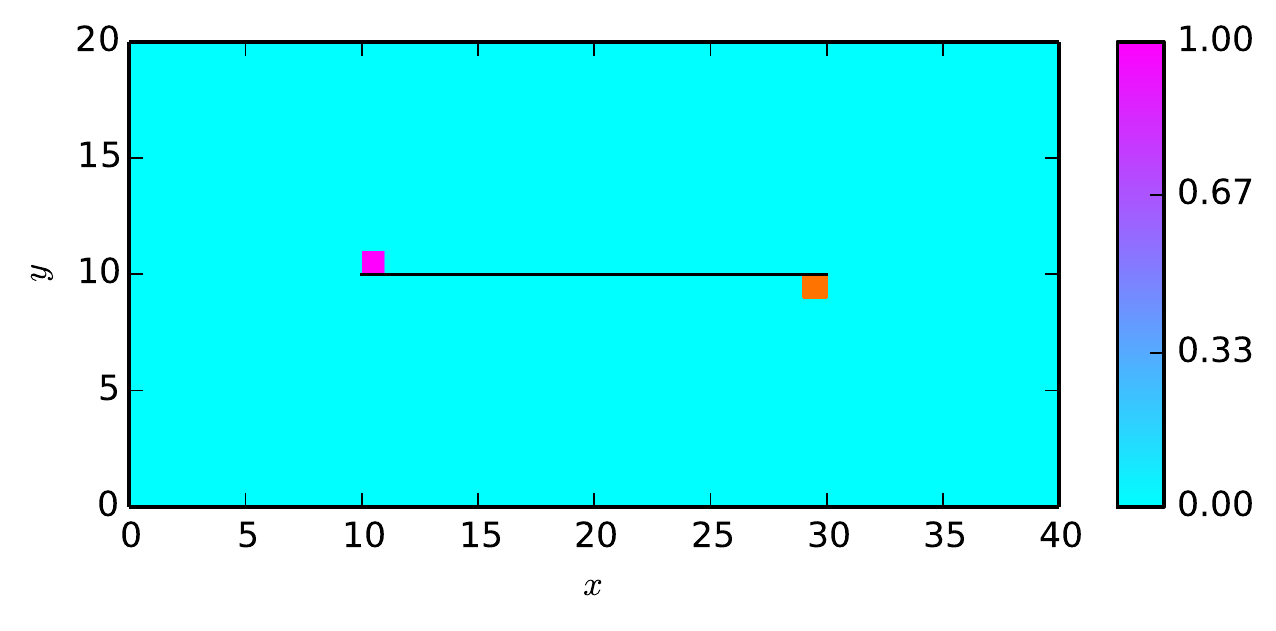}
  \includegraphics[width=\picwidth]{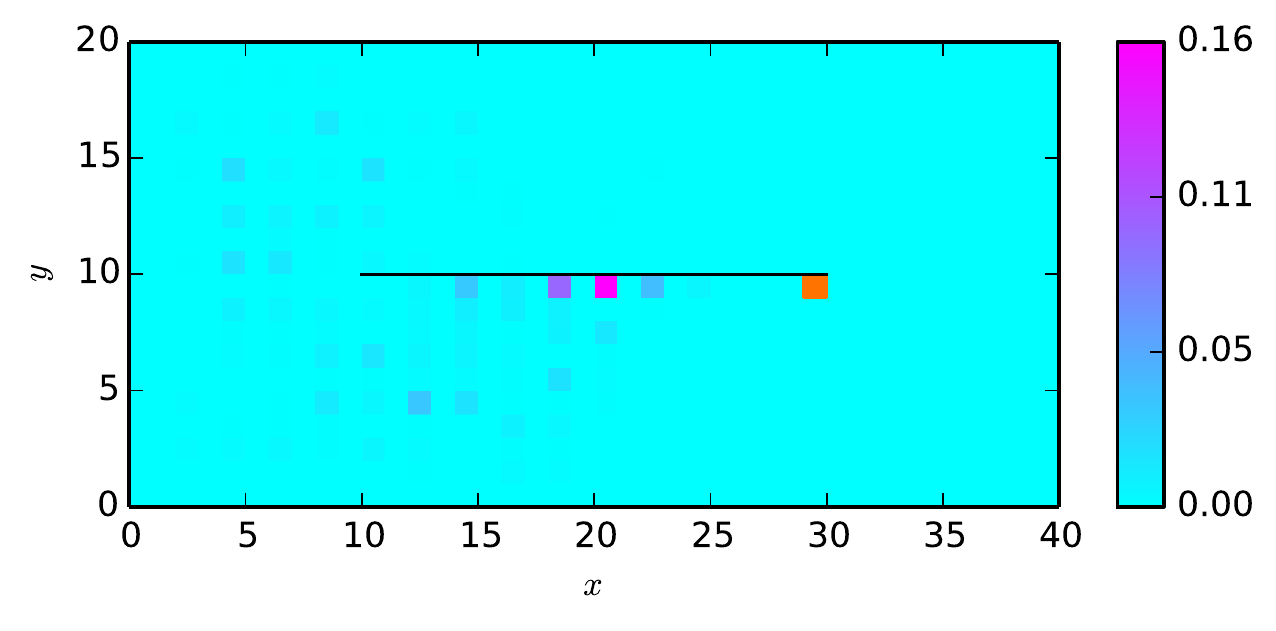}
  \includegraphics[width=\picwidth]{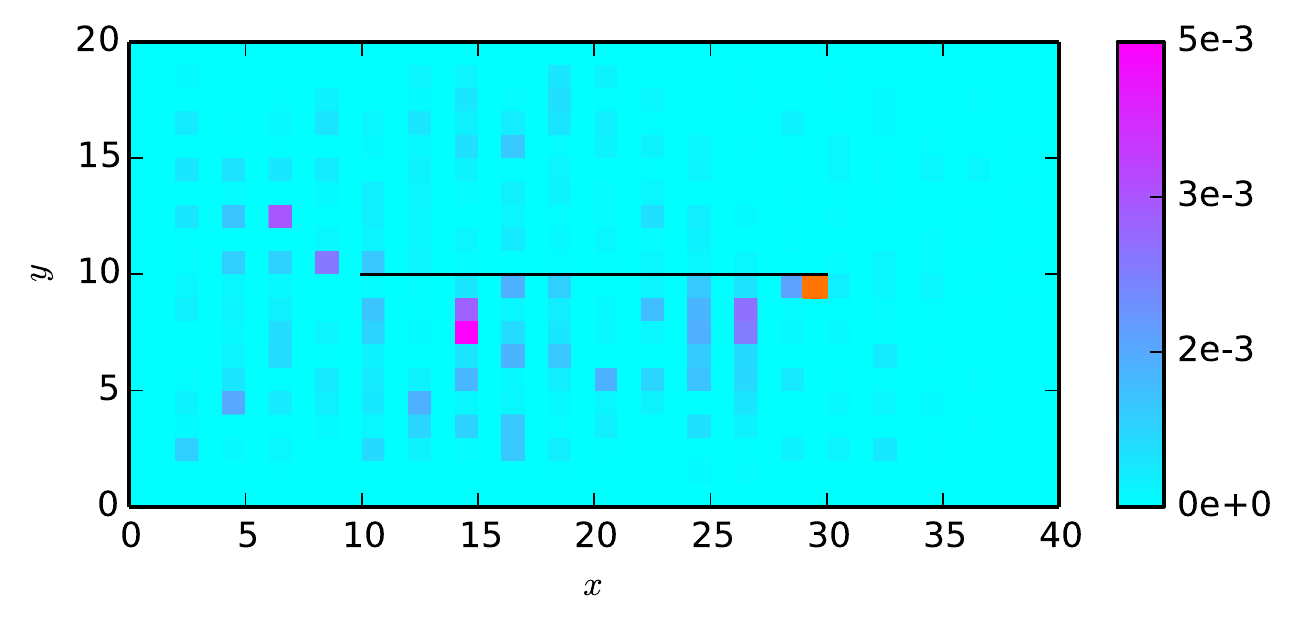}
  \caption{Wavefunction for a particular disorder realisation and the
    cut for a system size given by $M=10$. The values of rotation angles are $\theta_1=0.35\pi, \theta_2=0.15\pi, \delta=0.1\pi$. The cut is represented by the black line. The starting
    point for the wave function is just above the black line on the
    left. The final point at which the wave function gets absorbed is
    represented by the orange dot on the right-hand side. On the
    top  plot the wave function is plotted for $t = 0 $. The
    middle plot is for $t = 2M =20$ when a good fraction of the walker is
    on the conveyor. On the right-hand side the wave function is
    plotted for $t = 10M=100$, long after the bulk of the walker has been
    absorbed by the orange point B.}
  \label{fig:wavfn_cut_init}
\end{figure}

We choose a system of dimensions $(4 M\times 2 M) $. The walker is
initialised at the position $A=(M, M) $. The position of the final
(absorbing) point B is chosen to be $(3M-1, M-1) $. The cut cuts all
the links between the sites $(x, M) $ and $(x, M -1) $ for $M\leq
x\leq 3M$. Thus there is a path of cut links connecting the initial
and final site. For $M=10$ the system is plotted for three different
times in Fig.~\ref{fig:wavfn_cut_init}, thereby showing the initial
wavefunction, the wavefunction as it propagates along the conveyor and
the state after the majority of the wavefunction has been absorbed.
The boundaries of the system are absorbing boundaries.
This geometry is chosen
such that the walker cannot reach the absorbing boundary too quickly.

\begin{figure}[tb]
  \centering
  \def \picwidth {\columnwidth}
  \includegraphics[width=\picwidth]{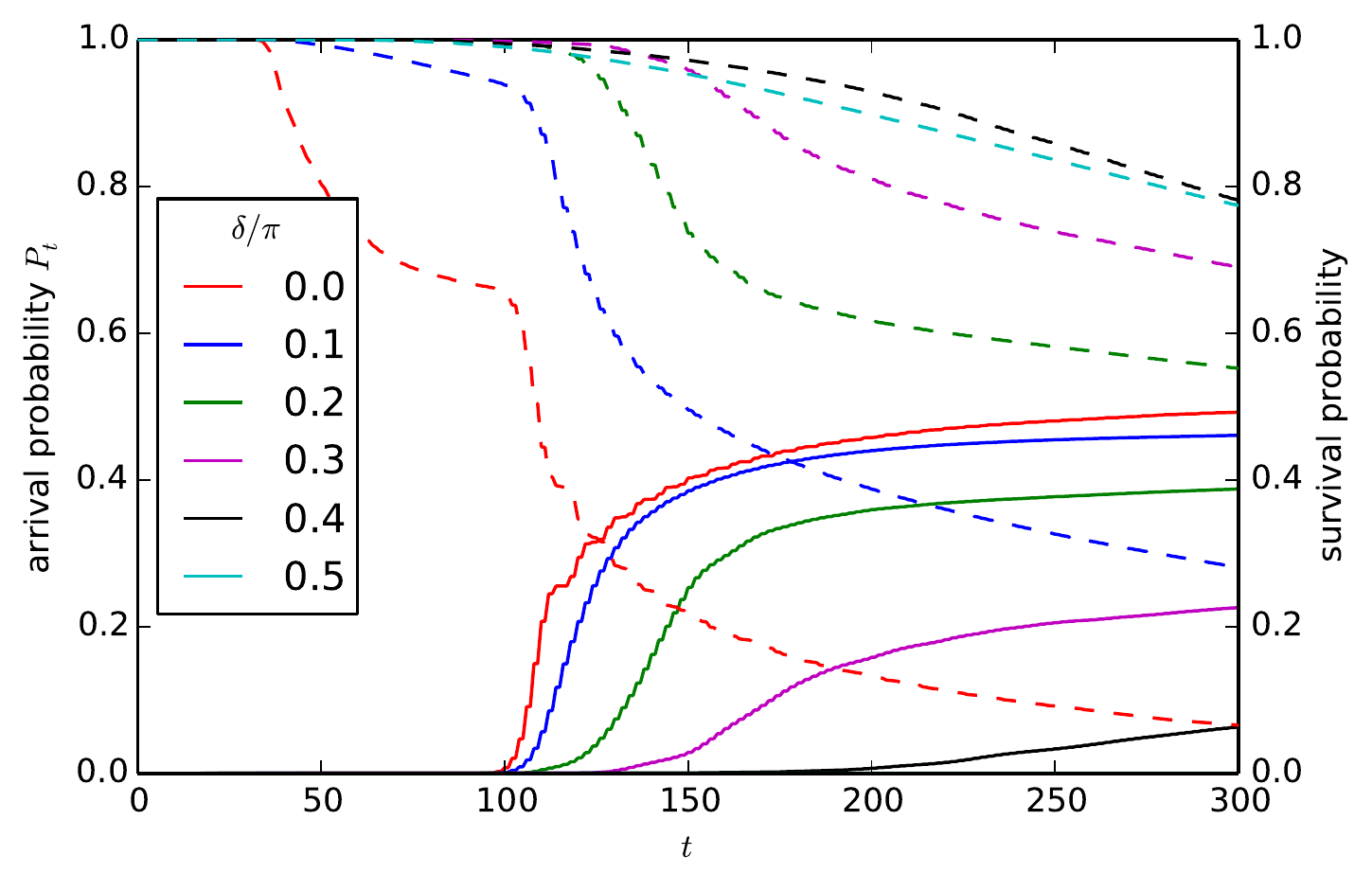}
  \includegraphics[width=\picwidth]{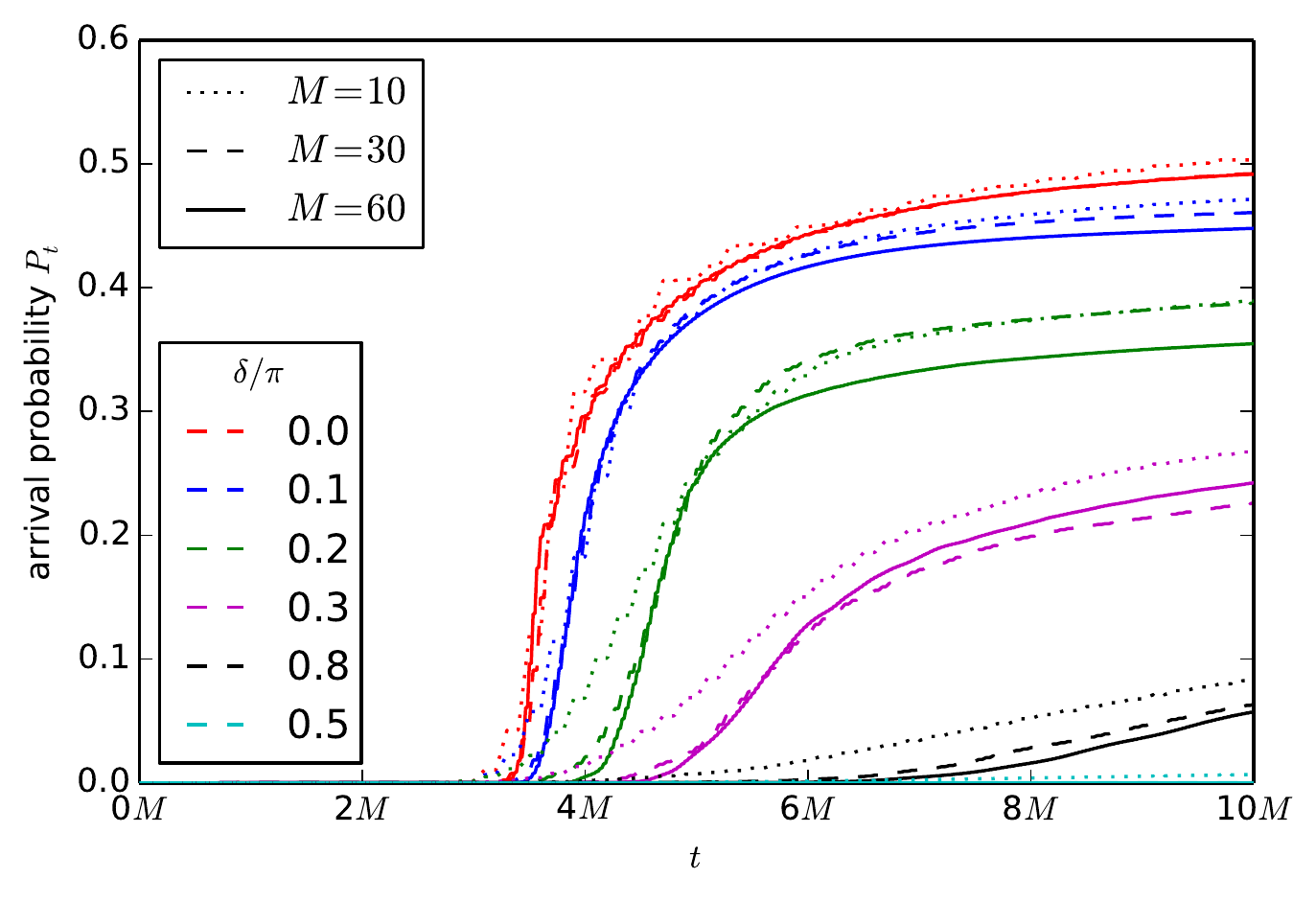}
  \caption{Top: Arrival and survival probabilities for
    $\theta_1=0.35\pi, \theta_2=0.15\pi$, $M=30$ and different amounts
    of rotation angle disorder $\delta$. Solid lines are cumulative
    arrival probabilities at the point B and dashed lines are the
    remaining wave function amplitudes, thus the probability of
    survival up to time $t $. The solid and dashed lines of the same
    colour correspond to the same system. We have averaged over 100
    different disorder realisations.  Bottom: Plot showing the arrival
    probability as a function of time for a different system sizes and
    different disorder strengths. The time axes is scaled with the
    system size. The curves for different system sizes collapse on one
    another, showing that the propagation along the cut is
    ballistic. The plot also shows that we may choose a system size of
    $M=30 $ in order to further investigate the system.}
  \label{fig:probs_vary_dtheta}
\end{figure}

We quantify the efficiency of the transport along the cut by looking
at the arrival probability $P_t$, as in
Eq. (\ref{eq:def_of_arrival_prob_Pt}) and the total survival
probability, i.e., the norm of the conditional wavefunction,
$\braket{\Psi(t)}{\Psi(t)}$. If these add up to 1, no part of the
walker is absorbed by the boundary.  If the walker is transported
ballistically along the defect we expect the total arrival probability
to suddenly increase by an appreciable amount at the time $ t=2M / v$,
where $v $ is the transport velocity of the walker, given in the clean
limit by eq. (\ref{eq:group_velocity}).  A delay in the onset of the
arrival at the final point B indicates a slowdown of the transport. On
the other hand, if the total survival probability decreases without
the probability at the final point B increasing, this also indicates a
loss of transport efficiency. It indicates that diffusion towards the
boundary increases in importance, whereas ballistic transport along
the cut decreases in importance. For different disorder strengths
$\delta$ we have plotted the results of such a calculation in
Fig.~\ref{fig:probs_vary_dtheta}.

One may obtain an overview of the behaviour as a function of
$\theta_1,\theta_2$ and $\delta$ by simply
looking at the total survival probability and the total arrival
probability for $t\gg 2M $ . Ballistically the walker should have arrived at the final
point B. This allows us to see whether the transport along a conveyor
is efficient for a range of parameters.
\begin{figure}[tb]
  \centering 
  \def \picwidth {\columnwidth}
  \includegraphics[width=\picwidth]{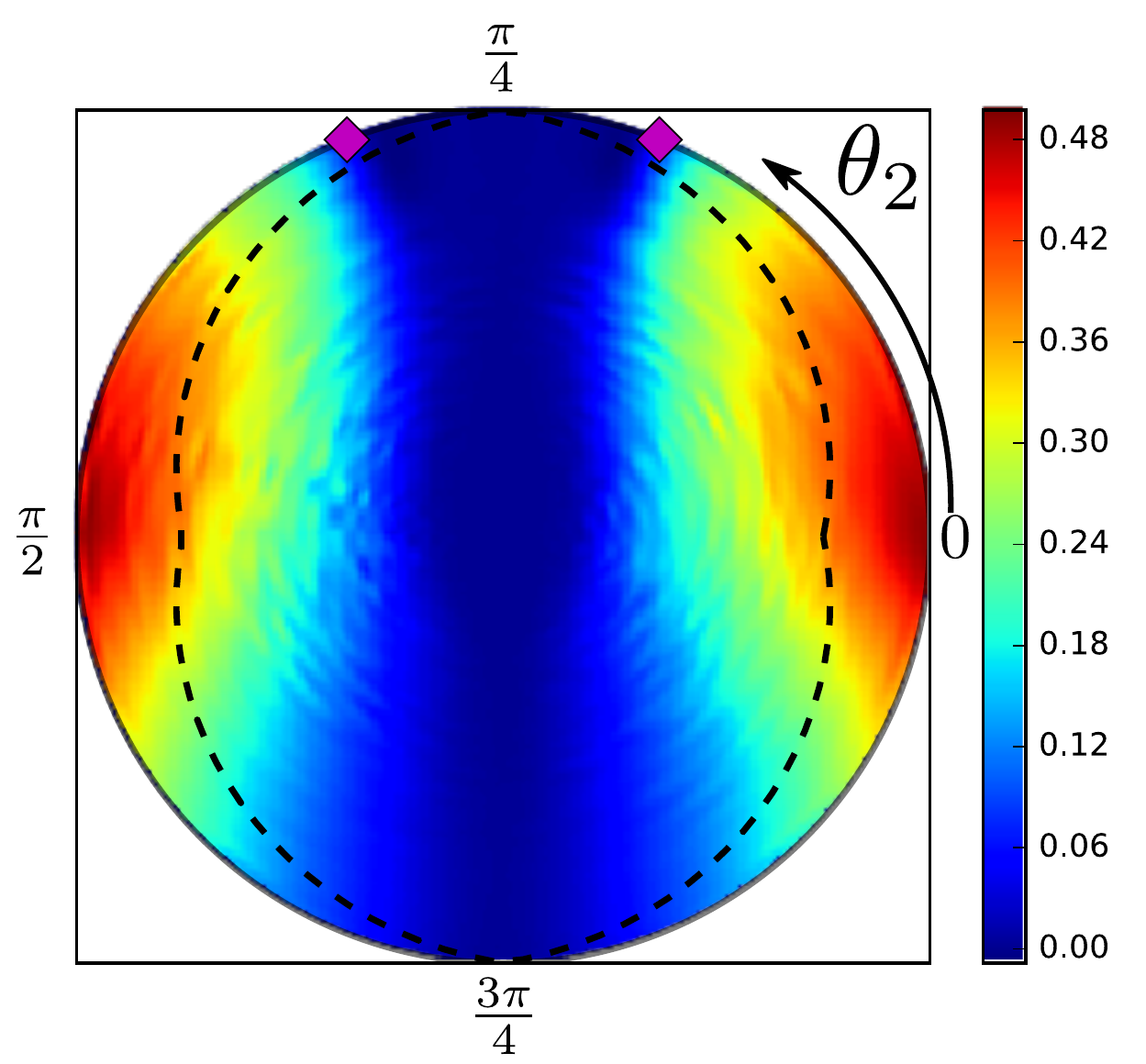}
  \caption{The arrival probability at point B after 322 time steps,
    averaged over 100 disorder configurations for $\theta_1=\frac\pi4$
    as a function of $\theta_2$ and $\delta$. The system has the same
    geometry as in Fig.~\ref{fig:wavfn_cut_init}, but is three times
    larger, having $M=30$. In the plot the azimuthal angle represents
    $\theta_2$ and the radius is related to $\delta$ by
    $r=1-2\delta/\pi$, such that the largest possible value of
    $\delta=\pi/2$ is taken at the centre at $r=0$, at which point
    $\theta_1$ and $\theta_2$ are irrelevant.  The black dashed line
    markes the regime at which $\delta$ becomes large enough for both
    types of topological invariants to be locally present in the
    system. Beyond that line transport begins to be suppressed.  The
    magenta diamonds mark the points at which the group velocity
    becomes too small for the walker to arrive within the simulation
    time. }
  \label{fig:surv_and_arr_prob_fn_th1_dth1}
\end{figure}

In Fig.~\ref{fig:surv_and_arr_prob_fn_th1_dth1} we have plotted the
final arrival probability for $\theta_1=\frac\pi4$, different values
of $\theta_2$ and a range of disorder strengths. We see that if
disorder is strong enough, the ballistic transport along the defect is
suppressed, and thus no part of the walker arrives at point B.
A naive expectation is that disorder can start to affect the edge
states only if it is large enough so that different
topological invariants can be present in different parts of the system.
This occurs for
\begin{align}
  \label{eq:criterion_of_delta_too_large}
  \delta>\delta_{max}=
  \begin{cases}
    \abs{\frac12(\theta_2-\pi/4)}  ,& \theta_2<\frac\pi2\\
    \abs{\frac12(3\pi/4-\theta_2} ,& \theta_2\ge\frac\pi2
  \end{cases}
\end{align}
The curve $\delta_{max}(\theta)$ is plotted as the dashed black line
in Fig.~\ref{fig:surv_and_arr_prob_fn_th1_dth1}: the numerical data 
are more or less in agreement with the naive expectation.  

The arrival probability also reduces to zero as $\theta_2$ approaches
$\theta_2=\frac\pi4$ and $\theta_2=\frac{3\pi}4$, independent of the
disorder. At the point $\theta_2=\frac\pi2$, we have 
$\sin(\theta_1-\theta_2)=0$ and thus the group velocity along the
conveyor is zero, cf. Eq. \ref{eq:group_velocity}. Since the walker
has to traverse a distance of $2M$ and the simulation time only runs
up to $t_{max}$, the walker will not arrive if
$v<v_{crit}=2M/t_{max}$. For
Fig.~\ref{fig:surv_and_arr_prob_fn_th1_dth1} $v_{crit}=0.19$. From
eq.~\eqref{eq:group_velocity} it then follows that the arrival
probability should be zero even in the clean limit when $\theta_2$ is
within a distance $\delta\theta_2^{crit}=0.06\pi$ of
$\theta_2=\frac\pi4$. These points are marked as magenta diamonds in
Fig.~\ref{fig:surv_and_arr_prob_fn_th1_dth1}. This estimate agrees
well with the position at which the arrival probability vanishes
in Fig.~\ref{fig:surv_and_arr_prob_fn_th1_dth1}.

Around the point
$\theta_2=\frac{3\pi}4$ on the other hand the group velocity does not
vanish. Instead, according to eq.~\eqref{eq:edge_penetration} the penetration depth of the edge state into the bulk
$\xi$ diverges. Thus
the overlap of the initial state of the quantum walk with the conveyor
vanishes, as initially the quantum walker is localised to a single
lattice site. Also the overlap of the conveyor state with the final
absorbing point disappears. Together with eq.~\eqref{eq:edge_penetration} this implies that the arrival 
probability $P_\infty$ around $\theta_2=\frac{3\pi}4$ will vanish as
\begin{align}
  P_\infty = 2 \,\delta\theta_z^2
  \label{eq:P_arr_around_34pi}
\end{align}
where $\delta\theta_z=\theta_2-\frac{3\pi}4$. We have numerically
checked this behaviour for the clean system and find that
eq.~\eqref{eq:P_arr_around_34pi} provides a good fit without any
adjustable parameters. So we observe qualitatively quite different
behaviour around the points $\theta_2=\frac\pi4$ and
$\theta_2=\frac{3\pi}4$. For $\theta_2=\frac\pi4$, $P_\infty$ vanishes
abruptly and stays zero over a finite range of $\theta_2$, namely, between the
two magenta diamonds in
Fig.~\ref{fig:surv_and_arr_prob_fn_th1_dth1}. On the other hand,
$P_\infty$ vanishes gradually around $\theta_2=\frac{3\pi}4$ and is
only strictly zero at one point.

\section{Conclusions}

In this work we have shown that in the 2-dimensional split-step
discrete-time quantum walk, a cut on the underlying lattice creates a
transport channel for the walker that is robust against
time-independent disorder. The mechanism for the transport is given by
edge states that form in the vicinity of the cut. We derived
analytical formulas for some properties of the edge states, and found
the bulk topological invariant that predicts their emergence. This
invariant is the winding of the quasienergy\cite{rudner_driven}.

The edge states we found are resistant to a moderate amount
time-independent disorder, but, as we have seen, above a certain
threshold they no longer exist. It is an interesting challenge to
study the details of this transition. In other words: how does
disorder destroy the topological phase? An important step in this
direction is understanding the effect of disorder on the 2DQW without
edges, our results on which are published
elsewhere\cite{jonathan_2014}.

There are quite promising perspectives for detecting the type of edge
states we found in quantum walk experiments. In fact, edge states due
to the Chern numbers have already been seen in a continuous-time
quantum walk experiment: there, the walker was a pulse of light
coupled into an array of waveguides etched into a block of dielectric,
a ``photonic topological insulator''\cite{rechtsman_photonic_2013}.
Modifying the pattern of the waveguides would allow for a direct
realization of the 2DQW. A more direct realization, which would also
allow the study of interactions, would be on ultracold atoms trapped
in an optical lattice\cite{alberti_electric_experiment}.

\begin{acknowledgments}
We acknowledge useful discussions with Mark Rudner, Carlo Beenakker
and Cosma Fulga. 
We also acknowledge the use of the
Leiden computing facilities. 
This research was realized in the frames of TAMOP
4.2.4. A/1-11-1-2012-0001 ''National Excellence Program -- Elaborating
and operating an inland student and researcher personal support
system'', subsidized by the European Union and co-financed by the
European Social Fund.  This work was also supported by the Hungarian
National Office for Research and Technology under the contract
ERC\_HU\_09 OPTOMECH, by the Hungarian Academy of Sciences (Lend\"ulet
Program, LP2011-016), and by by the Hungarian Scientific Research Fund
(OTKA) under Contract Nos. K83858 and NN109651.  
This work was funded by NORDITA.

\end{acknowledgments}

\bibliography{walkbib}{}

\appendix

\section{Propagation around a more complicated cut}
\label{app:star}

\begin{figure*}[tb]
  \centering
  \def \figw {.66\columnwidth}
  \includegraphics[width=\figw]{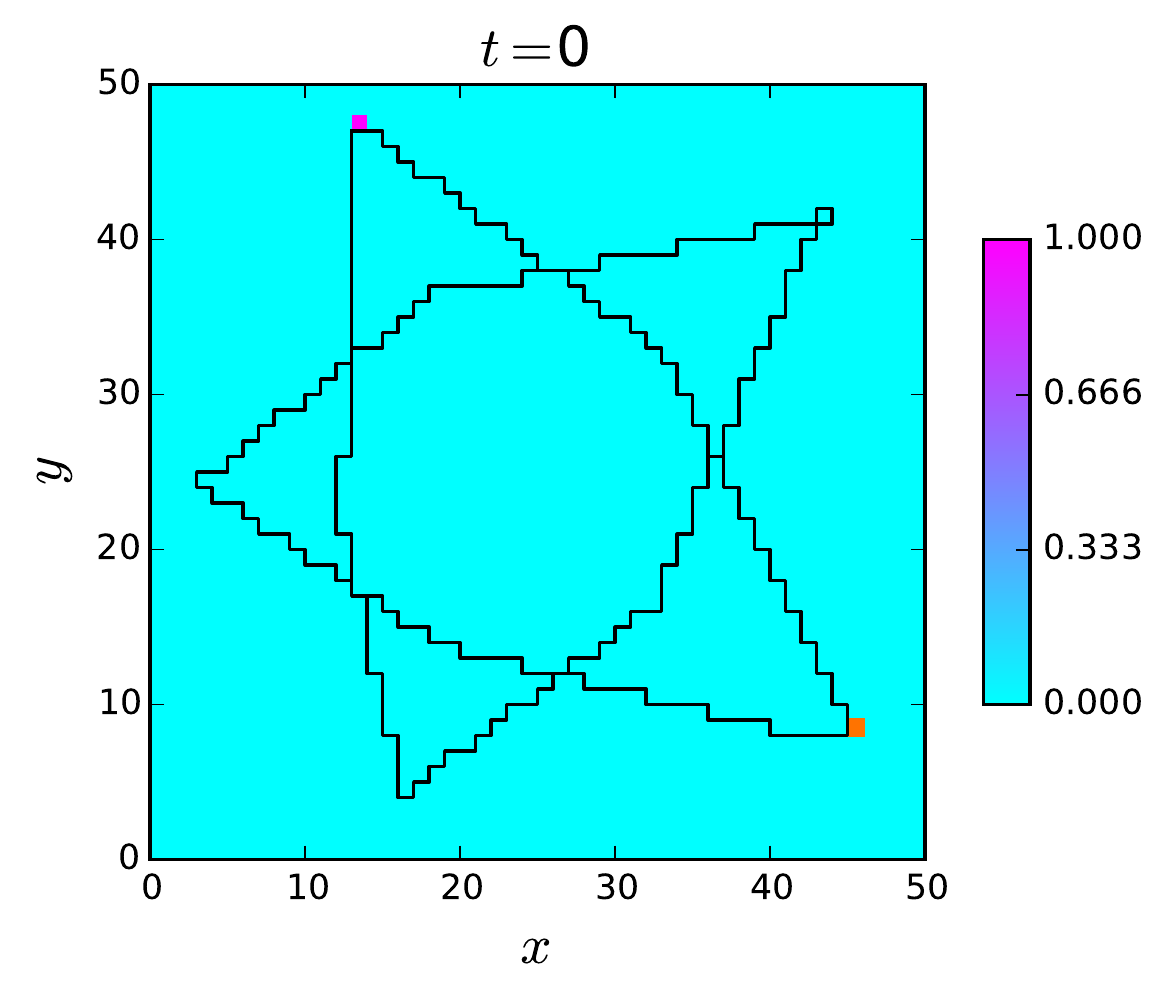}
  \includegraphics[width=\figw]{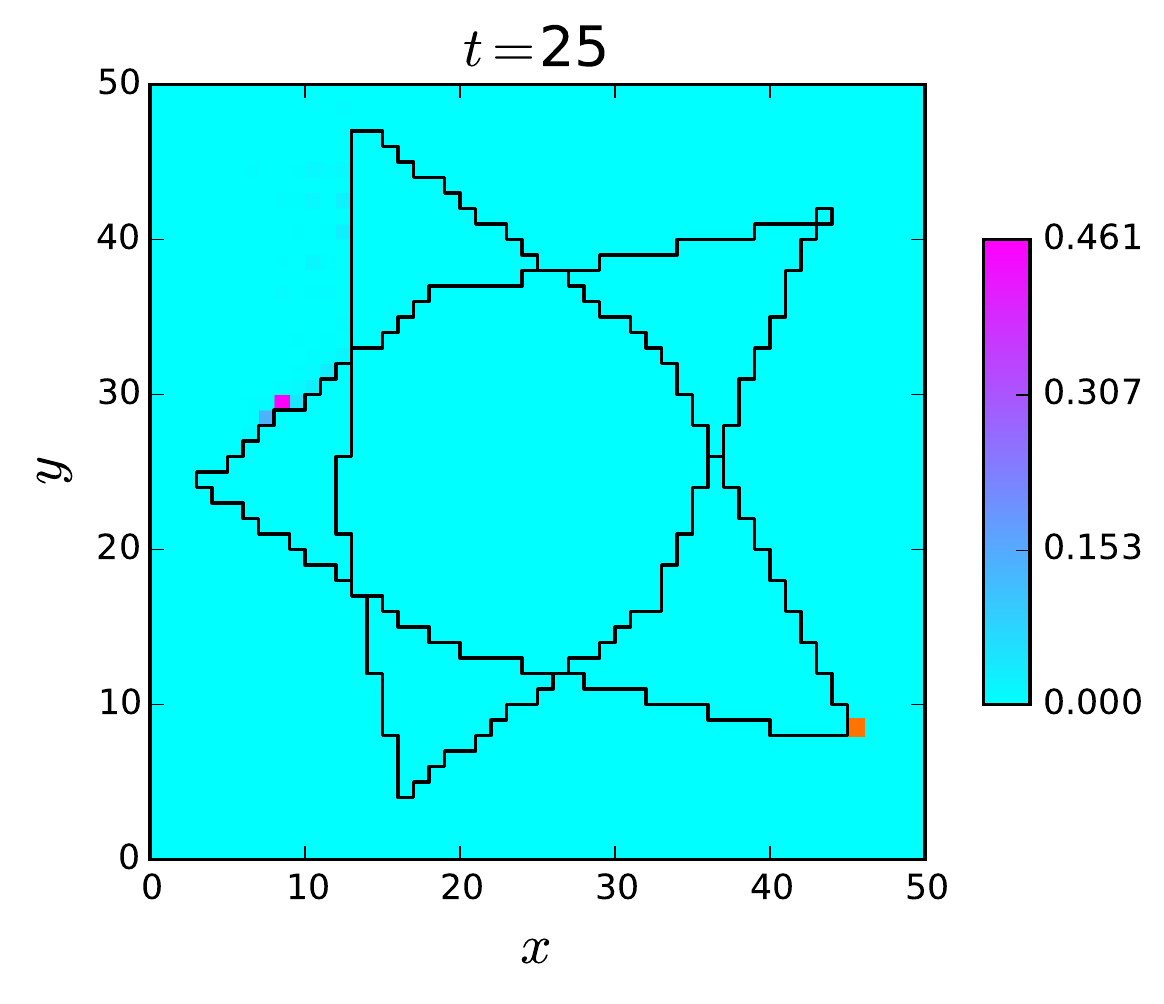}
  \includegraphics[width=\figw]{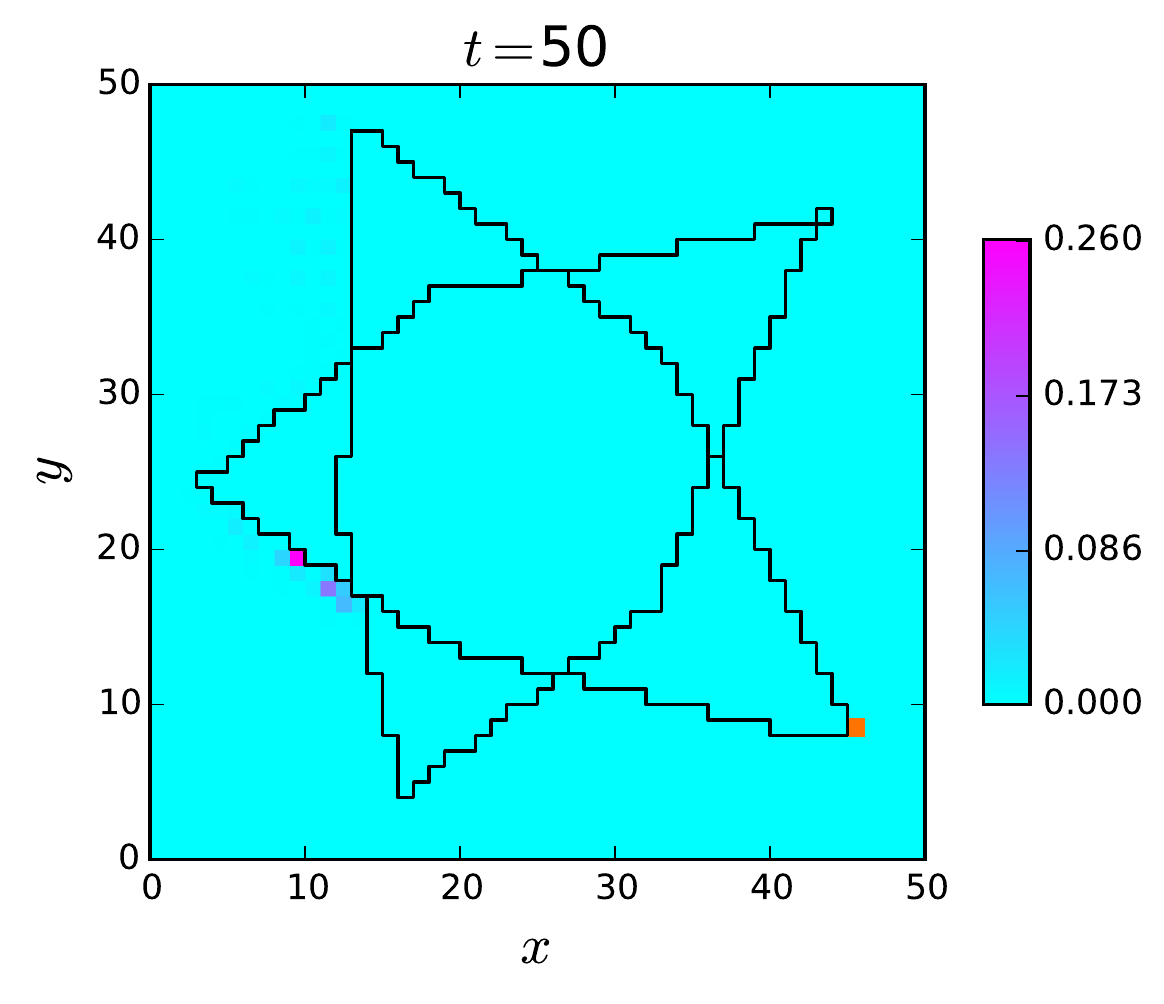}
  \includegraphics[width=\figw]{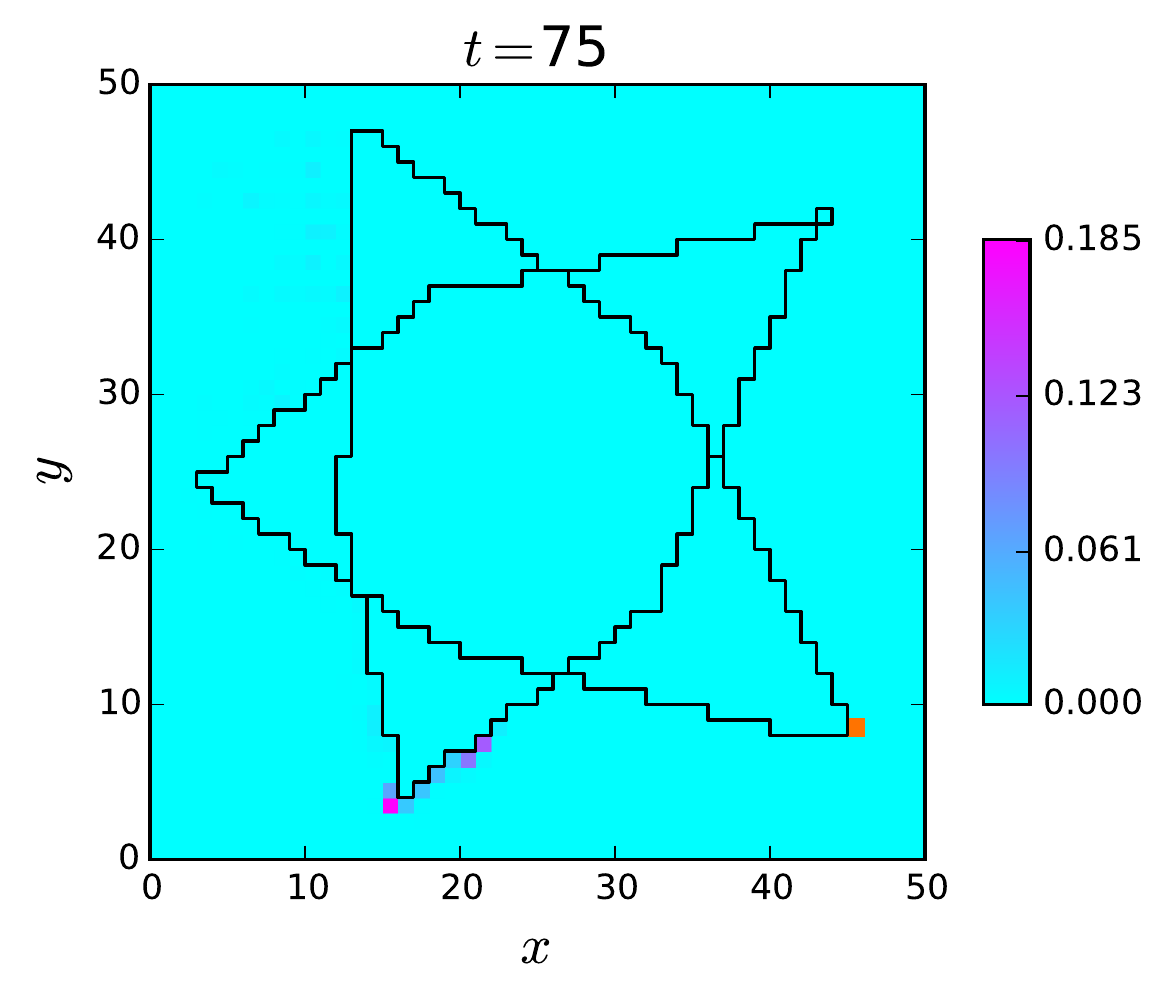}
  \includegraphics[width=\figw]{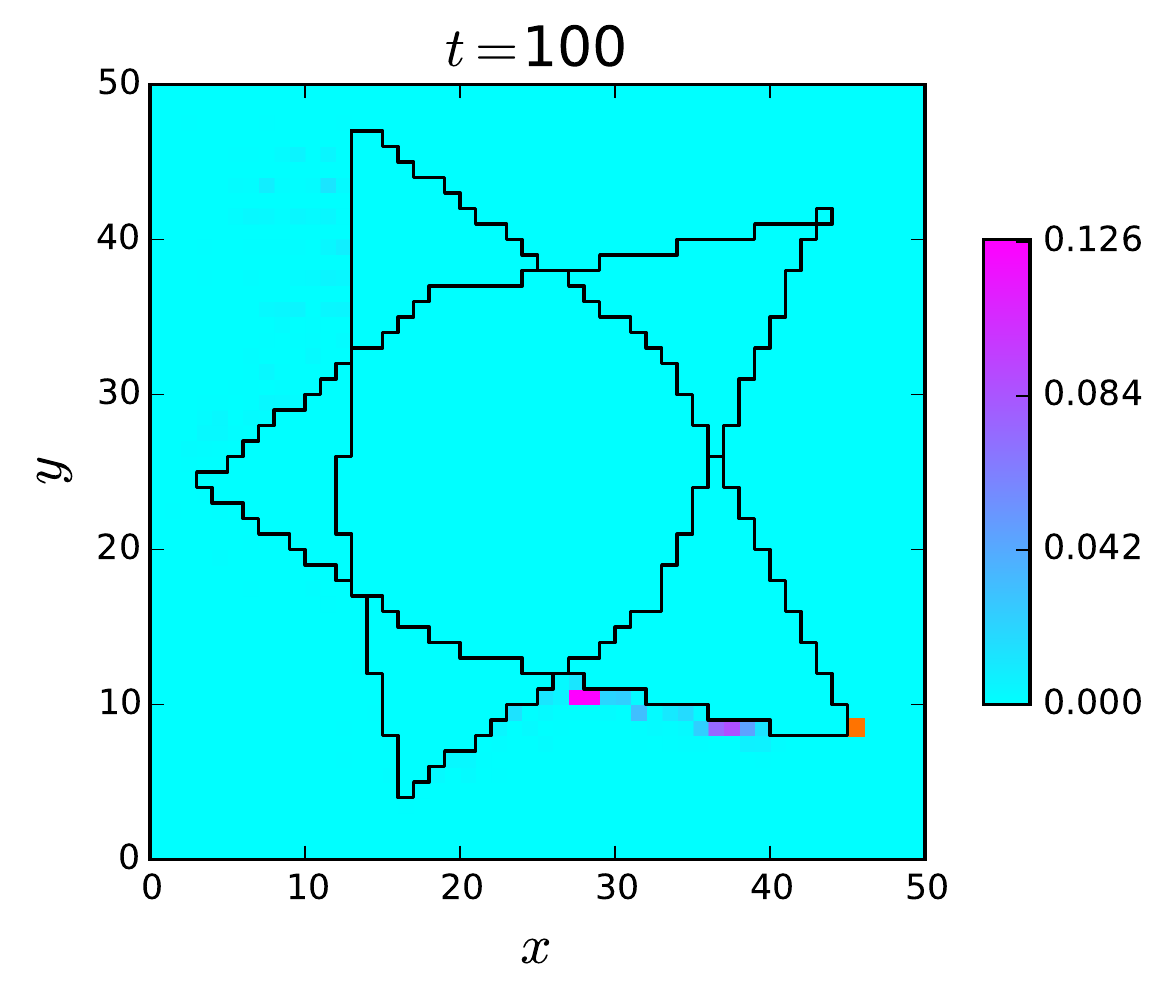}
  \includegraphics[width=\figw]{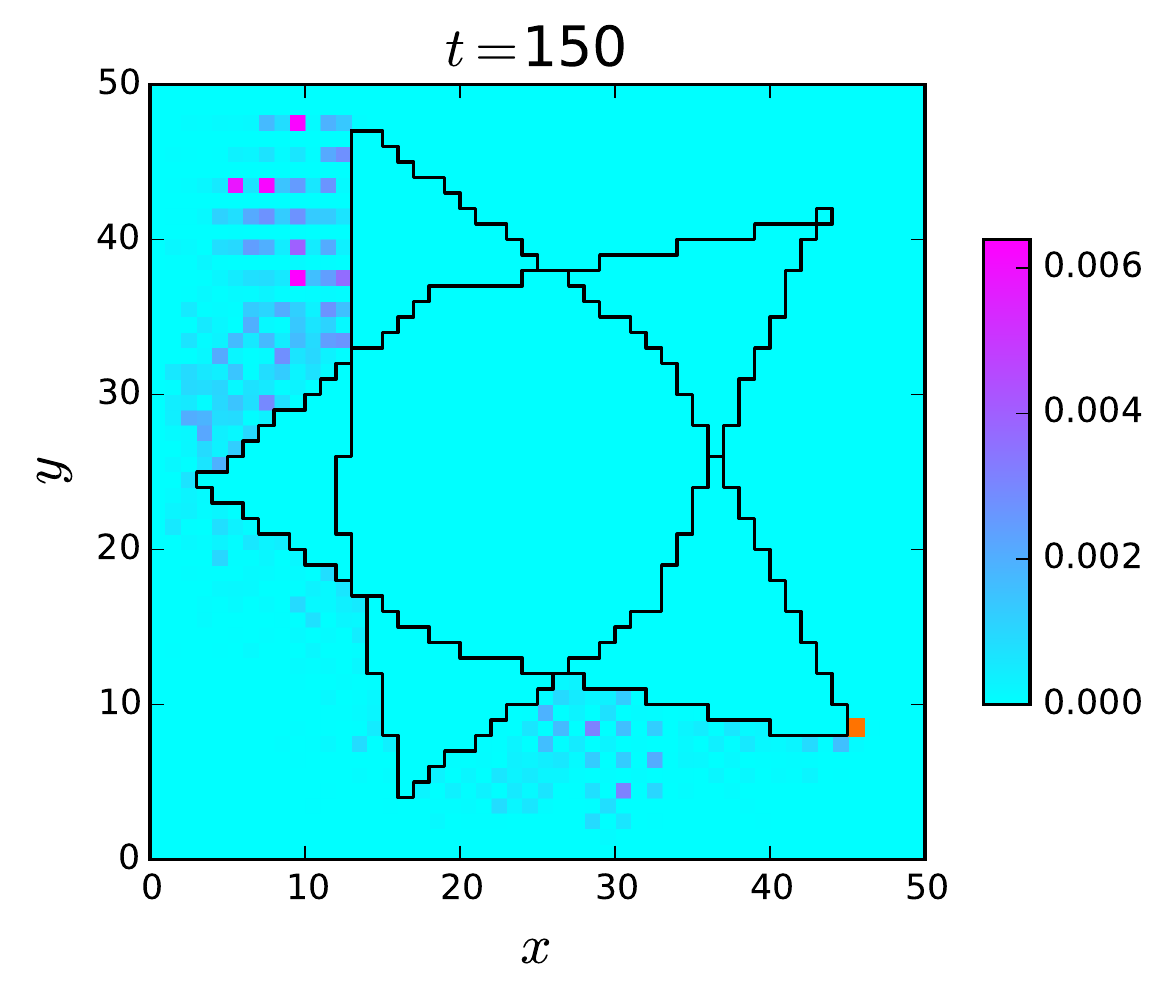}
  \includegraphics[width=\figw]{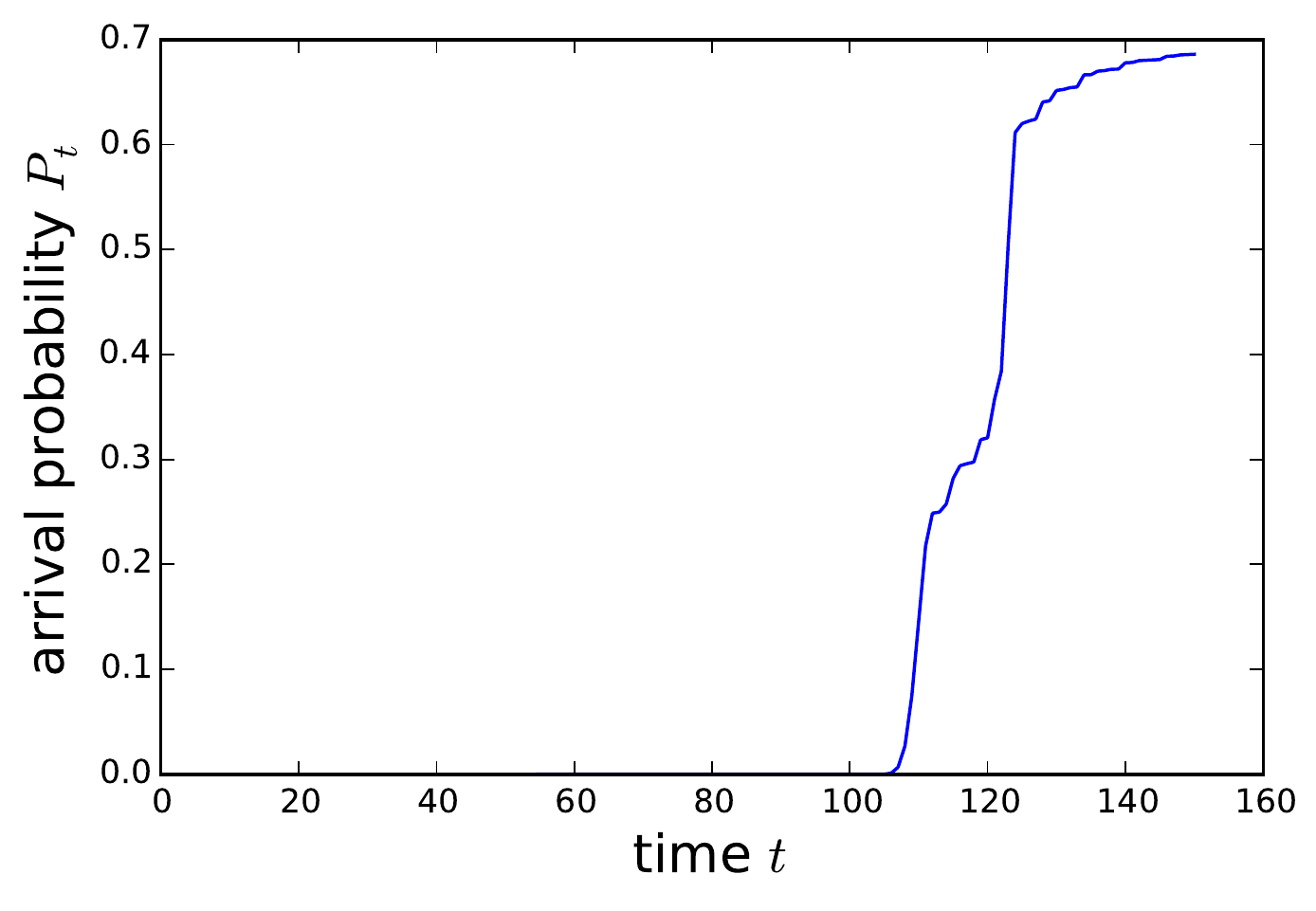}
  
  \caption{Illustration of the conveyor mechanism around a star-shaped figure for a single disorder realisation. We have plotted the wavefunction at different times $t$, starting at $t=0$ and ending at $t=150$, at which point the majority of the wavefunction amplitude has been absorbed by the final point, marked in orange. Note that the colour scale changes between the different panels. $\theta_1=0.45\pi, \theta_2=-0.05\pi$ and $\delta=0.1\pi$ were chosen, as, according to Eq.~\eqref{eq:group_velocity}, these maximise the propagation velocity along the cut. In the lowel panel we have plotted the arrival probability, which for large times approaches $P_t=0.7$.}
  \label{fig:conveyor_star}
\end{figure*}

In order to illustrate that the quantum walker can also follow a cut
which is not as simple as the one investigated in
section~\ref{sec:first_transport} and~\ref{sec:robustn-conv-belt}, we
have created a more complicated structure. This involves multiple
corners and also intersections of different cuts. In
Fig.~\ref{fig:conveyor_star} we investigate the propagation around a
star-shaped figure, choosing as a starting point one of the corners of
the star and as the endpoint another corner. We show the wave function
for 6 different time slices. We can clearly see that the quantum
walker propagates around the star.

\section{Edge State dispersion relations}
\label{app:edge_state}

In this Section we derive the edge state dispersion relations of edge
states of a 2DQW below a horizontal cut, using the transfer
matrix. We consider the 2DQW on a semi-infinite plane of integer
lattice points, i.e., $x,y\in\mathbb{Z}$ and $y<0$, with boundary
conditions given by the cut along $x$, above the line $y=0$. We assume
translation invariance along $x$, i.e., along the cut. In that case
the quasimomentum along $x$ is a good quantum number, we denote it by $k$.
Eigenstates
of the walk can be taken in a plane wave form, 
\begin{align}
\Psi_k(x,y,s) = e^{ikx} \Psi_y^s, 
\end{align}
with $s = \uparrow$ or $s=\downarrow$.
Since the shift along $x$ can be written as $S_x = e^{-ik \sigma_z}$, 
the eigenvalue equation of the walk reads 
\begin{align}
U &= S_y V(k);\\
V(k) &= e^{-i\theta_2 \sigma_y}e^{-ik \sigma_z} e^{-i\theta_1 \sigma_y}.
\end{align}
Note that apart from $ \text{det } V = 1$, we also have 
\begin{align}
{V^{\downarrow \downarrow}}^\ast &= V^{\uparrow \uparrow} = c_+ \cos k
- i c_- \sin k;\\
 V^{\downarrow \uparrow} &= -{V^{\uparrow
    \downarrow}}^\ast = s_+ \cos k +i s_- \sin k,
\end{align}
where we use the shorthands
\begin{align}
s_\pm &= \sin(\theta_1 \pm \theta_2);\\
c_\pm &= \cos(\theta_1 \pm \theta_2).
\end{align}

The boundary conditions on the edge states for $y\to-\infty$ is that
their wavefunctions should be normalizable. To put this into an equation, 
we first find a 
suitably defined transfer matrix.  We consider an eigenstate
$\ket{\Psi}$ of $U$ with quasienergy $\varepsilon$, for whose
components we have
\begin{subequations}
\begin{align}
e^{-i\varepsilon} \Psi_{n+1}^\uparrow &= V^{\uparrow\uparrow} \Psi_n^\uparrow + 
V^{\uparrow\downarrow} \Psi_n^\downarrow;
\\
e^{-i\varepsilon} \Psi_{n-1}^\downarrow &= V^{\downarrow\uparrow} \Psi_n^\uparrow + 
V^{\downarrow\downarrow} \Psi_n^\downarrow.
\end{align}
\label{eq:eigval_M}
\end{subequations}
The transfer matrix is defined by 
\begin{align}
\begin{pmatrix}
\Psi_{y+1}^\uparrow\\\Psi_{y}^\downarrow
\end{pmatrix} &= 
M(\varepsilon)  
\begin{pmatrix}
\Psi_{y}^\uparrow\\\Psi_{y-1}^\downarrow
\end{pmatrix}.
\end{align}
Substituting into \eqref{eq:eigval_M} gives us 
\begin{align} 
M(\varepsilon) &= \frac{
\cos \varepsilon + i \sin \varepsilon \sigma_z 
+ \text{Re}V^{\uparrow\downarrow} \sigma_x 
- \text{Im}V^{\uparrow\downarrow} \sigma_y}
{V^{\downarrow \downarrow}} .
\end{align}
For the eigenstate $\ket{\Psi}$ to be normalizable, 
the vector
$(\Psi_{-1}^{\uparrow}, \Psi_{-2}^{\downarrow})$ must be an eigenvector of the
transfer matrix $M$ with eigenvalue whose absolute value is higher
than one.  
The eigenvalues of $M$ are 
\begin{align}
m_\pm &= 
\frac{\cos \varepsilon \pm \sqrt{\abs{V^{\uparrow\downarrow}}^2-\sin^2\varepsilon}}{V^{\downarrow \downarrow}}.
\label{eg:m_eigen}
\end{align}

If $\pm\cos \varepsilon > 0$, the normalizable edge state corresponds to
the eigenvalue of the transfer matrix $m_\pm$, 
and we need 
\begin{align}
\begin{pmatrix}
\Psi_{-1}^\uparrow \\
\Psi_{-2}^\downarrow
\end{pmatrix}
 &\propto 
\begin{pmatrix}
V^{\uparrow \downarrow} \\ \pm \sqrt{\abs{V^{\uparrow\downarrow}}^2-\sin^2\varepsilon} - i \sin \varepsilon
\end{pmatrix}.
\label{eq:y_to_infty}
\end{align}


We next consider the boundary condition on the top of the ribbon, $y=0$. 
Here, because of the cut link,
%
realized by $-i\sigma_y$, we have
\begin{align}
\Psi_{0}^\downarrow e^{-i\varepsilon} &= \left(V^{\uparrow \uparrow} 
\Psi_{0}^\uparrow + V^{ \uparrow \downarrow} \Psi_0^\downarrow \right).
\end{align}
This is easiest to solve if we choose 
\begin{align}
\Psi_{0}^\downarrow &= V^{\uparrow \uparrow};&
\Psi_{0}^\uparrow &= e^{-i\varepsilon} - V^{\uparrow\downarrow}.
\label{eq:reflect1}
\end{align}
Using Eqs.~\eqref{eq:eigval_M}, we obtain 
\begin{align}
\Psi_{-1}^\downarrow &= 
\label{eq:reflect2}
e^{i\varepsilon} + V^{\downarrow\uparrow}.
\end{align}

Combining the two boundary conditions, Eq.~\eqref{eq:y_to_infty} with
Eq.~\eqref{eq:reflect1} and Eq.~\eqref{eq:reflect2} above, we have
\begin{align}
\frac{e^{-i\varepsilon} + {V^{\downarrow\uparrow}}^\ast}
{e^{i\varepsilon} + V^{ \downarrow\uparrow}}
=
\frac{-{V^{\downarrow \uparrow}}^\ast}
{\pm \sqrt{\abs{V^{\downarrow\uparrow}}^2-\sin^2\varepsilon} - i \sin \varepsilon}
,
\label{eq:eigen_M}
\end{align}
for $\pm\cos \varepsilon>0$.
The absolute values of the left-and the right-hand-side of this
equation are both 1, so this is really an equation for the phases.


\subsubsection{The gap around quasienergy $\varepsilon=0$}

Consider $\varepsilon=0$; then Eq.~\eqref{eq:eigen_M} reads
\begin{align}
\frac{1+\alpha^\ast}{1+\alpha} &= -\frac{\alpha^\ast}{\abs{\alpha}},
\end{align}
with 
\begin{align}
\alpha &= s_+ \cos k +i s_- \sin k.
\end{align}
Solving this equation for $\text{arg}(\alpha)$, we obtain $\text{arg}\alpha=\pi$, which implies 
\begin{align}
k = \pi &\text{ if } s_+>0;\\
k = 0 &\text{ if } s_+<0.
\end{align}

The edge state wavefunctions decay exponentially towards the bulk, as
$\abs{\Psi_y^s} = \abs{\Psi_0^s} e^{-\abs{y}/\xi}$. To obtain their
penetration length for $\pm s_+ >0$, we substitute $\cos
\varepsilon=1$, $\sin k=0$, $\cos k=\pm 1$, into
Eq.~\eqref{eg:m_eigen}, and get
\begin{align}
\xi &=-\left( \text{log } \frac{1-\abs{s_+}}{\abs{c_+}} \right)^{-1}.
\end{align}

To obtain the group velocities, we solve Eq.~\eqref{eq:eigen_M} around
$\varepsilon\approx 0$.  If $s_+>0$ ($s_+<0$), then we have $k\approx
0$ ($k\approx \pi$).  In both cases, if we use $k$ to denote the small
distance from 0 or $\pi$, we find to 1st order in the small parameters
$\varepsilon$ and $k$,
\begin{align}
V^{\uparrow \downarrow} &= \mp s_+ \pm ik s_-,\\
\sqrt{\abs{V^{\uparrow \downarrow}}^2-\varepsilon^2} &= \pm s_+.
\end{align}
So Eq.~\eqref{eq:eigen_M} transforms to 
\begin{align}
\frac{1+ i\varepsilon \mp s_+ \pm ik s_-}{1- i\varepsilon \mp s_+ \mp ik s_-}
&=
\frac{ \mp s_+ \pm ik s_-}{ \mp s_+ - i\varepsilon}. 
\end{align}
To first order in the small parameters, this gives us 
\begin{align}
\frac{\varepsilon}{k} = \mp s_- 
\end{align}

\subsubsection{The gap around quasienergy $\varepsilon=\pi$}
The edge states in the quasienergy gap around $\varepsilon=\pi$ are
the sublattice partners of the edge states around $\varepsilon=0$.
Due to the sublattice symmetry of the quantum walk, any eigenstate of
the walk at quasienergy $\varepsilon$ with wavefunction $\Psi(x,y)$
has a sublattice partner with quasienergy $\varepsilon+\pi$ and
wavefunction
\begin{align}
\Gamma \Psi(x,y) = e^{i\pi x} e^{i\pi y} \Psi(x,y).
\label{eq:gamma_shift_momentum}
\end{align}
Thus, for a fixed value of the rotation angle parameters $\theta_1$
and $\theta_2$, edge states in the gap at $\varepsilon\approx \pi$
have the same penetration depth and group velocity as those at 
$\varepsilon \approx 0$, and are around $k=\pi$ ($k=0$) when those in the gap 
around $\varepsilon=0$ are around $k=0$ ($k=\pi$). 

\subsubsection{Summary: group velocity and penetration depth}

To summarize, at the middle of both of the gaps around $\varepsilon=0$
and $\varepsilon=\pi$, the edge states have the group velocity
\begin{align}
\frac{d \varepsilon}{d k} &= 
\text{sign}(\theta_1+\theta_2) \sin(\theta_2-\theta_1),
\end{align}
and penetration depth
\begin{align}
\xi &=-\left( \text{log } 
\frac{1-\abs{\sin(\theta_1+\theta_2}}{\abs{\cos(\theta_1-\theta_2)}} 
\right)^{-1}.
\end{align}

\section{Sublattice symmetry of a Quantum Walk and Chern numbers}
\label{app:sublattice_symmetry}

To understand the sublattice symmetry of the split-step quantum walk,
assign each site on the lattice one of four sublattice indices,
\begin{align}
f(x,y) = 2(y \text{ mod } 2) + (x + y) \text{ mod } 2, 
\end{align}
and use the corresponding sublattice projection operators, 
\begin{align}
\Pi_j = \sum_{x,y: f(x,y)=j} \ket{x,y}\bra{x,y},
\end{align}
where $j \in \{0,1,2,3 \}$.  One timestep $U$ changes the sublattice
index by $2$, as can be checked explicitly. Thus, a walker started at
$x_0,y_0$ on sublattice $j$ will be on sublattice $j+2 \text{ mod } 2$
after an odd number of timesteps, and return to sublattice $j$ after
an even number of timesteps. Now define the sublattice operator
$\Gamma$ as
\begin{align}
\Gamma &= \Pi_0+\Pi_2 - \Pi_1 -\Pi_3.
\end{align}
This operator acts on a wavefunction $\Psi(x,y)$ as 
\begin{align}
\Gamma \Psi(x,y) = e^{i\pi x} e^{i\pi y} \Psi(x,y).
\label{eq:gamma_shift_momentum}
\end{align}
When acting on a plane wave, $\Gamma$ shifts its wavenumber by
$(\pi,\pi)$. On the other hand, acting on an eigenstate of the walk,
it shifts the quasienergy by $\pi$, since
\begin{align}
\Gamma U \Gamma &= -U& \to \quad \Gamma H_\text{eff} \Gamma = H_\text{eff}+\pi. 
\end{align}
This means that every band with Chern number $C$ has a sublattice
symmetric partner that is shifted in energy by $\pi$ with the same
Chern number $C$. Since the sum of all Chern numbers have to be 0, in
a two-band model, such as the split-step walk, this precludes the
existence of a band with a nonzero Chern number.

\end{document}